\documentclass[twocolumn,showpacs,preprintnumbers,amsmath,amssymb,nofootinbib]{revtex4-1}

\usepackage{graphicx}
\usepackage{dcolumn}
\usepackage{bm}
\usepackage{verbatim} 
\usepackage{epsfig}
\usepackage{amsmath}
\usepackage{amsthm}
\usepackage{subfigure}

\begin{document}

\date{\today}

\title{ETEAPOT: symplectic orbit/spin tracking code for all-electric 
storage rings}
\author{Richard M Talman}
\affiliation{
Laboratory of Elementary Particle Physics,
Cornell University, Ithaca, NY, USA}
   
\author{John D Talman}
\affiliation{
UAL Consultants, Ithaca, NY, USA}

\begin{abstract}
Proposed methods for measuring the electric dipole moment (EDM) of the
proton use an intense, polarized proton beam stored in an 
all-electric storage ring ``trap''. At the ``magic'' kinetic energy of 
232.792 MeV, proton spins are ``frozen'', for example always parallel 
to the instantaneous particle momentum. Energy deviation from the magic 
value causes \emph{in-plane} precession of the spin relative to the momentum. 
Any non-zero EDM value will cause \emph{out-of-plane} precession---measuring
this precession is the basis for the EDM determination. A proposed 
implementation of this measurement shows that a proton EDM value of 
$10^{-29}\,$e-cm or greater will produce a statistically-significant, measurable 
precession after multiply-repeated runs, assuming small beam depolarization 
during 1000 second runs\cite{pEDM-PRL}, with high enough precision to 
test models of the early universe developed to account for the present day 
particle/anti-particle population imbalance.

This paper describes an accelerator simulation code, ETEAPOT, a new
component of the Unified Accelerator Libraries (UAL), to be used
for long term tracking of particle orbits and spins in electric 
bend accelerators, in order to simulate EDM storage ring experiments. 
Though qualitatively much like magnetic rings, the non-constant particle
velocity in electric rings give them significantly different properties, 
especially in weak focusing rings. Like the earlier code TEAPOT 
(for magnetic ring simulation) this code 
performs \emph{exact tracking in an idealized (approximate) lattice} rather than the more 
conventional approach, which is \emph{approximate tracking in a more nearly exact lattice.} 
The BMT equation describing the evolution of spin vectors through idealized 
bend elements is also solved exactly---original to this paper. Furthermore 
the idealization permits the code 
to be exactly symplectic (with no artificial ``symplectification''). 
Any residual spurious damping or 
anti-damping is sufficiently small to permit reliable tracking for 
the long times, such as the 1000 seconds assumed in estimating the 
achievable EDM precision.

This paper documents in detail the theoretical formulation implemented 
in ETEAPOT. The accompanying paper\cite{AGSAnalogue} 
``EDM planning using ETEAPOT with a 
resurrected AGS Electron Analogue ring'' describes the practical 
application of the ETEAPOT code in the UAL environment to ``resurrect'', 
or reverse-engineer, the ``AGS-Analogue'' all-electric ring built at 
Brookhaven National Laboratory in 1954. Of the (very few) all-electric 
rings ever commissioned, the AGS-Analogue ring is the only relativistic 
one and is the closest to what is needed for measuring 
proton (or, even more so, electron) EDM's. That paper also describes
preliminary lattice studies for the planned proton EDM storage 
rings as well as testing the code against analytic calculations.

\end{abstract}


\maketitle


\section{Introduction}

\paragraph{\bf Orbit and spin simulation code needed for electric storage rings.}
The U.S. particle physics community has recently updated its vision of the future
and strategy for the next decade in a Particle Physics Project Prioritization 
Panel (P5) Report. One of the physics goals endorsed by P5 is measuring 
the EDM of fundamental particles (in particular proton, deuteron, neutron and 
electron).

Since Standard Model EDM predictions are much smaller than current experimental 
sensitivities, detection of any particle's non-zero EDM would 
signal discovery of New Physics. If of sufficient strength, such a source could 
provide an explanation for the observed matter/antimatter asymmetry of our 
universe. A proton EDM collaboration\cite{pEDM-PRL} has 
proposed a storage ring proton EDM measurement at the unprecedented level of 
$10^{-29}e \cdot\,$cm, an advance by nearly 5 orders of magnitude beyond the 
current indirect bound obtained using Hg atoms.

This paper is limited to the theoretical orbit and spin dynamical formulation 
within ETEAPOT, which is a newly developed
code within the Universal Accelerator Libraries (UAL) simulation
environment.
The accompanying paper ``Using ETEAPOT to resurrect the AGS-Analogue 
ring for EDM planning'' describes the practical application of the
ETEAPOT code with emphasis on details of simulation requirements for
the EDM measurement. 

\paragraph{\bf Complications imposed by electric bending.}
The fundamental complication of an electric ring, as contrasted 
to a magnetic ring, is the non-constancy of particle speed. 
A fast/slow separation into betatron and synchrotron amplitudes 
has become fundamental to the conventional (Courant-Snyder) 
magnetic ring formalism. But, in an electric lattice the 
mechanical energy (as quantified by the relativistic 
factor $\gamma=1/\sqrt{1-\beta^2}$) varies on the same time 
scale as the transverse $x$ and $y$ amplitudes.
On the other hand, changing only in RF cavities, the 
total energy $\mathcal{E}=\gamma mc^2+eV(r)$, which includes
also the potential energy $eV(r)$, changes on a slow time
scale, which makes a similar fast/slow 
separation valid. 

In the magnetic formalism, since it is only in RF cavities 
that the mechanical energy varies, the $\gamma$ factor is 
invariant in the rest of the ring. Furthermore one is 
accustomed to treating energy as constant for times 
short compared to the synchrotron period. To the extent 
the betatron parameters are independent of particle energy, 
the betatron and synchrotron motions can then be superimposed.

To most closely mimic this treatment in an electric ring,
and to continue to regard $\gamma$ as the fundamental
``energy-like'' parameter, requires us to evaluate 
$\gamma$ only in regions of zero electric potential,
which is to say, not in RF cavities, and not in electric
bending elements---in other words, only in field free
drift regions. This leads to a curious, but entirely
satisfactory, representation in which the particles spend
most of their time inside bend elements where $\gamma$ is
varying, and little time in short drift regions where
$\gamma$ is constant. The reason this is fully satisfactory
is that the drift regions are closely spaced, and more or 
less uniformly spaced around the ring. Knowing the lattice
functions exactly at those points is operationally
equivalent to knowing them everywhere.
With this interpretation, one can again
rely on the approximate representation of the motion
as a superposition of fast betatron and slow synchrotron motions.

\section{Particle tracking paradigms}
The conventional formulation of accelerator physics is based on a paraxial 
approximation in which all orbit angles are small relative to a central design 
orbit. Not only is this approximation quite good for small rings, it becomes
progressively better as ring radii increase. The most important formulas 
obtained in this approach are
based on linearization of the paraxial orbit equations. For sufficiently small
amplitudes orbit evolution can be represented by ``transfer matrices'' multiplying
initial displacements, represented in 6D phase space by six component displacement
vectors. By introducing nonlinear ``transfer maps'' this approach can be extended
to larger amplitudes by representing each output variable as a (truncated) 
power series of terms each of which is a product of initial 
components, or their squares, cubes, etc. This approach can be referred to as 
\emph{approximate tracking in an ``exact'' lattice} 
where ``exact'' means, for example, that magnetic field profiles, fringe 
fields etc., once represented faithfully by accurate power series, can be 
subsumed into the previously described truncated power series orbit representations.

The TEAPOT tracking paradigm\cite{TEAPOT} is very different. It is a refinement
of the ``kick code'' paradigm, under development when TEAPOT was first 
introduced, about thirty years ago. Though still based on the paraxial 
approximation, elements are 
sliced into sufficiently short segments that they can be represented by 
delta function ``kicks'', or kinks, where orbit slopes change 
discontinuously, but displacements 
are continuous. The main virtue of this approach is that it is symplectic; 
i.e. particle beams respect Liouville's theorem.

TEAPOT took the more extreme approach of jettisoning the paraxial approximation
altogether, and insisting that orbits be constructed only from exactly 
determined, analytic orbits, interupted where appropriate, by kicks. This
can be referred to as \emph{exact tracking in an approximate lattice}. This approach is 
restricted by the fact that exact orbits are known only in very
special cases; for example in uniform magnetic fields. However, by introducing
thin kick elements  it is possible to represent actual elements and actual 
rings to good accuracy. These added kicks may be artificial, meaning 
they model field deviations from the idealized field, or they may represent
the total fields of elements physically present in the lattice.
All this is
explained in detail in the original Schachinger and Talman paper\cite{TEAPOT},
which remains valid, and in use, to this day. To restrict the length of the 
present paper, only features newly introduced in ETEAPOT, and not covered
in the original paper, are described. 

One new feature is the requirement to use electric rather than magnetic bending 
elements, which brings in the complication already mentioned, that, 
unlike in magnets, 
particle speeds are not constant in electric elements. The other new feature is 
the need to also track the particle spins. To respect the TEAPOT paradigm this
further restricts the list of allowable lattice elements to elements for which
the BMT equation, an abbreviation for Michel, Bargmann, and 
Telegdi\cite{BMT}, is exactly solvable. Fortunately it is possible, 
even straightforward, to meet both of these new restrictions.

The ETEAPOT approach has been developed to address new requirements of the proton 
EDM project. First is the transition from mgnetic to electric bending. 
Magnetic tracking codes like TEAPOT and MAD implicitly take advantage of the constant 
speed of particles in magnetic fields.  This seriously complicates the porting of 
simulation algorithms to electric rings in which the particle speed depends on the 
local electric potential energy. Thick bending elements are again to be replaced by 
integrable force fields. As it happens, Inverse square law is the unique  
force field for which the 3D relativistic equations of motion can be solved exactly. 
The second  major ETEAPOT extension has been the introduction and tracking of 
particle spin vectors (which provides the EDM signature of the EDM experiment). 
The differential equation  governing spin motion is the BMT equation. In general 
3D motion the BMT equations are computationally challenging. But in our 
idealized elements the BMT equations can be solved exactly, as required by 
the fundamental TEAPOT/ETEAPOT design principle. 

By symmetry, every orbit in a radial force field lies in a plane, referred to
here as ``the bend plane''. This plane is always close to but, in general,
not quite identical to the horizontal symmetry of the ring. 
In spherical $(r,\theta)$ coordinates the differential equation governing 
$r(\theta)$ for a relativistic particle orbit in an inverse square law force field
can be solved exactly\cite{Munoz}. These are like classical planetary orbits or 
the orbits in a hydrogen atom treated classically. Even relativistically,
every orbit is a precessing ellipse-like (rosette) figure, 
familiar to Newton in classical mechanics, and for relativistic motion to Einstein, 
in his general relativistic calculation of the advance of the perihelion of Mercury.  

Using $(r,\theta)$ polar coordinates, it will be shown shortly that orbit evolution 
in an inverse square law bend plane is described exactly by
\begin{equation}
r(\theta) = \frac{\lambda}{1+\epsilon\cos\kappa(\theta-\theta_0)},
\quad\hbox{and}\quad
x = r - r_0
\label{eq:Munoz.14m}
\end{equation}
where $r$ is the radial distance from the bend center
to the particle position. Then $x$ is the (paraxial) radial displacement 
from the design central orbit, which is a circle of radius $r_0$. 
This formula is as simple as it is only because the 
motion is two dimensional. The potential energy depends only on
$r(\theta)$, permitting the speed, and hence the momentum components to 
be expressed analytically as functions of $\theta$. In correlating this 
equation with conventional paraxial treatment, one notes that the independent 
variable $\theta$ here differs from the independent longitudinal variable $s$, 
which is the path length along the design orbit; but $s$ and $\theta$ 
advance proportionally in bend elements, with the constant of proportionality 
being the design radius $r_0$. 

Each particle being tracked has its own private orbit plane parameters 
in the orbit equation---though they are all very nearly the same for realistic
beams. The parameters in Eq.~(\ref{eq:Munoz.14m}) are interpretable (approximately) 
as follows: $\lambda$ is ``average'' radius, $\epsilon$ is ``eccentricity'', 
$\theta_0$ is the angular deviation from perihelion, and $\kappa$ 
(interpretable as the ``tune'' in accelerator jargon) establishes the
betatron phase advance per unit angular advance.

For particle tracking this formulation is simpler than the 3D description of 
orbits in a uniform magnetic field. One price to be paid for this simplicity is 
the need to transform each particle into, and later, out of, its private 2D 
orbit plane (with its own basis vectors)
as it enters, and later, leaves a bending element. 
These transformations, not shown here, are elementary near-identity rigid 
rotations (including spin). 

A more serious price for this exact, ``global'' coordinate, approach 
is purely numerical. Though Eq.~(\ref{eq:Munoz.14m}) is 
analytically exact it is numerically treacherous---the paraxial quantity $x$
is a millimeter scale length compared to $r$ which is very nearly equal to 
the $r_0$ which is a large length, such as 40\,m. The special numerical treatment
needed to handle this issue is discussed below.

\section{Relativistic kinematics in electric potential $V(r)$.}
\begin{figure}[ht]
\centering
\includegraphics[scale=0.4]{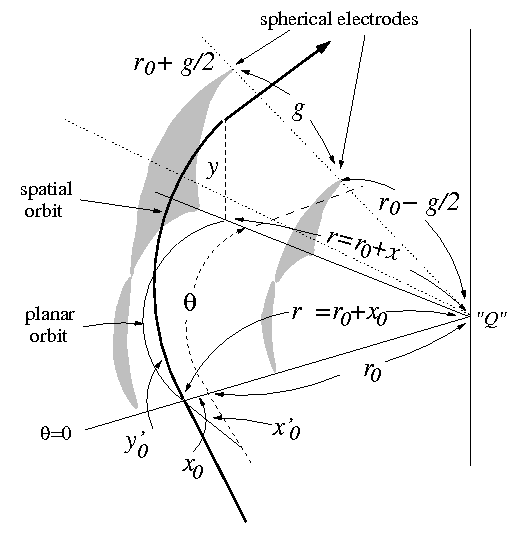}
\caption{\label{fig:SphericalElectrodes}
The bold curve shows a particle orbit passing through a 
\emph{spherical}, $m=1$, electrostatic bending element. The 
shaded surfaces are electrodes and the figure is
grossly distorted. The ``$Q$'' shown at the origin is the
``effective point charge'' that would give the same
idealized electric field as the electrodes.}
\end{figure}

In the horizontal $y$=0 bend plane, a radial electric field with index $m$ 
power law dependence on radius $r$ is
\begin{equation}
{\bf E}(r,0)
 = 
-E_0\,\frac{r_0^{1+ m }}{r^{1+ m }}\,{\bf\hat r},
\label{eq:WeakFoc.1m}
\end{equation}
and the electric potential $V(r)$, adjusted to vanish 
at $r=r_0$, is 
\begin{equation}
V(r)
 =
-\frac{E_0r_0}{ m }\,
\bigg(
\frac{r_0^m }{r^m }
 -
1
\bigg).
\label{eq:WeakFoc.2m}
\end{equation}
The ``cleanest'' case, shown in Figure~\ref{fig:SphericalElectrodes}, 
has $m$=1 and is known
as the Kepler or the Coulomb electric field of a point source.
One way our case is more general than Kepler's is that our
treatment has to be exactly relativistic. This $m$=1 case can
be referred to as ``spherical'' since the
equipotentials are spheres centered on a point charge
at the center,
and $r$ is the radius in $(r,\theta,\phi)$ spherical coordinates.
For $m$=1 the Kepler problem can be solved in closed form with the
same generality as in the relativistic as in the nonrelativistic 
case. However the orbits are no longer exactly elliptical, nor 
restricted to a single plane.

\begin{figure}[ht]
\centering
\includegraphics[scale=0.35]{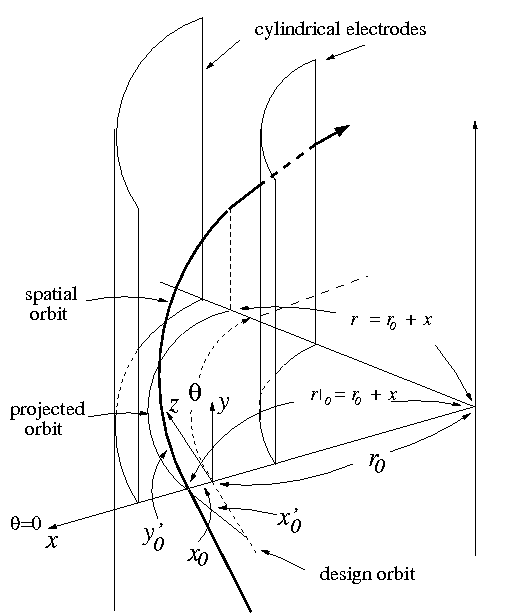}
\caption{\label{fig:CylindricalElectrodes}
The bold curve shows a particle orbit passing through a
curved-planar \emph{cylindrical}, $m=0$, electrostatic bending element. 
The electrode spacing is $g$ and the design orbit
is centered between the electrodes in the $y$=0 plane.}
\end{figure}

The $m$=0 field is referred to as ``cylindrical'' and $r$
is the radius in $(r,\phi,y)$ cylindrical coordinates. 
This field is the experimentally easiest to produce, using 
cylindical electrodes having the same central axis. 
It has been tacitly assumed that the EDM storage ring bends
will be constructed this way, even if the optimal electrode shape
would be somewhat different. As already explained, (weak) thin 
quadrupoles can be used to tailor the fields to overcome
this impediment. Curved-planar electrodes produce
the electric field shown in 
Figure~\ref{fig:CylindricalElectrodes}

\paragraph{\bf Solution of the equation of motion.}
Throughout much of this section formulas of Mu\~noz and
Pavic\cite{Munoz} will be transcribed essentially unchanged, 
except for bringing symbols into consistency with conventional
accelerator notation. The Mu\~noz/Pavic formulation, though equivalent 
to various other formalisms describing relativistic Coulomb
orbits, is especially appropriate for our relativistic
accelerator application. Mu\~noz and Pavic
show that the ``generalized''-Hamilton vector
\begin{equation}
\widetilde{\bf h}
 = 
\widetilde{h_r}\,{\bf\hat r} + \widetilde{h_{\theta}}\,{\hat{\bm\theta}}
\label{eq:Munoz.1}
\end{equation}
is especially useful for describing 2D relativistic Kepler
orbits. Our 3D accelerator application can be formulated in such a way
as to use only such 2D orbits. Except for scale factors, and a 
$\gamma$-dependent offset, the pair $(\tilde h_r, \tilde h_{\theta})$ will reduce
to the conventional phase space coordinates $(x,x')$.\footnote{The overhead 
tildes in Eq.~(\ref{eq:Munoz.1}) have been added for
our later convenience, when, for our accelerator application,
we will rescale the generalized-Hamilton vector 
slightly, for example making it dimensionless rather than a velocity.
Dividing by a constant characteristic velocity will bring the
formalism more into conformity with standard accelerator terminology.
Except for these overhead tildes, in this section, quantities and
formulas will be largely copied from M \& P; this includes
using the abbreviation $f'$ for $df/d\theta$. When the formulas
have finally been interpreted in accelerator context, the usual
accelerator definition of the prime symbol, with $f'$ standing
for $df/ds$, where $s$ is arc length, will be adopted. On the 
design orbit of radius $r_0$, where $\theta$ and $s$ are both defined,
one has $s=r_0\theta$, and this rescaling merely introduces
constant factors $r_0$ into the formulas.}

In accelerator context the
evolution of $\widetilde{\bf h}$ for any particular particle
will be interpretable as the phase space evolution 
of that particle.
We will refer to $\widetilde{\bf h}$ as
the ``MP-vector'' since, as far as we know, Mu\~noz and
Pavic intoduced it. In the non-relativistic regime 
$\widetilde{\bf h}$ is related to
the (nonrelativistically conserved) Laplace-Runge-Lenz vector. 
As generalized
by Mu\~{n}oz and Pavic, though not quite conserved in the
relativistic regime, $\widetilde{\bf h}$ satisfies a very simple 
equation of motion, whose exact solutions are sinusoids, even
at arbitrarily large ``betatron'' amplitudes.

In geometric mechanics jargon\cite{RT-GM}, the lattice is therefore 
``integrable''. The $\widetilde {\bf h}$ formalism then makes it
possible to define an implicit transfer matrix (where ``implicit''
implies the matrix elements depend on each particle's 
own conserved orbit parameters, a.k.a. ``orbit elements''). 
This has the virtue of hiding the strong coupling between kinetic energy 
and horizontal position in an electric ring.

Though $\widetilde {\bf h}$ is not conserved in general, 
Mu\~noz and Pavic show (and it will soon become obvious)
that \emph{$\widetilde {\bf h}$ is conserved if and only if the orbit is circular.} 
For the proton EDM lattice the design orbit is circular within
bends. Also, for weak-focusing lattices, neglecting the effects of 
lattice quadrupoles situated between the bends, off-momentum closed 
orbits are also almost circlular. 

But $\widetilde {\bf h}$ also has the numerical
disadvantage mentioned earlier---transverse displacements (the normally
significant aspect of accelerator dynamics) need,
at least superficially, to be
described by expressions having large cancellations. 
This makes it essential to use series 
approximations very carefully, even when the 
linearized paraxial approximation is fully valid.  
It is important to organize the
evolution formulas in such a way as to exploit
cancellations analytically, for example arranging for
leading terms to cancel analytically, rather than relying 
on their numerical cancellation. This can usually be 
accomplished (without sacrificing exact expressions)
by simplifying the sum of the approximately cancelling terms
and keeping track and restoring any discrepancy.
This recovers numerical precision for 
small amplitudes comparable to that achieved in conventional 
Courant-Snyder formulation.

Another inconvenient aspect of the Mu\~noz/Pavic formalism
is the fact that the design, central orbit is
\emph{degenerate} in the sense that it is, in fact, 
the arc of a perfect circle. This
is precisely the condition in which the perihelion
(or aphelion) angular orbit element is undefined because the
radius vector length is independent of angular position.
This ambiguity is most important in the time of flight 
calculations needed for accurate modeling of synchrotron
(longitudinal) oscillations---a calculation that strains the 
numerical accuracy because all path lengths are so nearly equal.

These same considerations complicate
the derivation of explicit (meaning the elements are 
independent of particle coordinates) 
transfer matrices and their generalization
to truncated power series transfer maps.
This also is only a inconvenience that can be
worked around. 

With qualitative discussion complete, the analytic formulation
can begin. The Lorentz force equation in the $m$=1 spherical case is
\begin{equation}
\frac{d{\bf p}}{dt}
 =
- k\,\frac{{\bf\hat r}}{r^2},
\label{eq:Munoz.2}
\end{equation}
where $k$ is the customary MKS notation for 
$1/(4\pi\epsilon_0)$ except for implicitly including also the
(point) charge factor. On the central orbit the centripetal force 
equation is
\begin{equation}
\frac{(p_0c/e)\beta_0}{r_0}
 =
\frac{k/e}{r_0^2}
 \overset{\rm also}{\ =\ }
E_0.
\label{eq:Munoz.2p}
\end{equation}
where $k$, like $E_0$, the on-axis electric field, is defined to 
be positive. In UAL, as in the MAD simulation code, energies like $p_0c$ 
are always expressed numerically in GeV units, such that their numerical 
value in the computer is $10^{-9}p_0c/e$, giving them GV units---an MKS 
unit. From Eq.~(\ref{eq:Munoz.2p}), 
$k/e$=$(p_0c/e)\beta_0r_0$ and $k$ is always expressed in the
computer as $10^{-9}k/e$, which gives it GV-m units---also an
MKS unit. The on-axis electric field $E_0$ is measured in GV/m units. 
In these units $k=E_0r_0^2$, numerically, with $r_0$ measured in meters.

A particle's angular momentum is 
\begin{equation}
{\bf L}
 =
{\bf r}\times{\bf p}.
\label{eq:Munoz.3}
\end{equation}
In terms of the design momentum $p_0$ and the design
radius $r_0$, $L_0=r_0p_0$, and
\begin{equation}
\frac{k}{L_0c} = \beta_0\quad ( = 0.598379\ {\rm for\ pEDM}).
\label{eq:Munoz.3pm}
\end{equation}
Like $k$, internally (i.e. in the computer), $L_0c$ is expressed 
numerically as
$10^{-9}L_0c/e$. For internal numerical values to be most easily 
substituted into an analytic external documentation
formula, factors should be grouped 
as $pc/e$, $Lc/e$, $k/e$, etc. or, as here, $k/(Lc)$, and factors of
$10^9$ (for eV to GeV conversion) put in ``by hand''.
Eq.~(\ref{eq:Munoz.3pm}) is also useful in the form
\begin{equation}
k = \gamma_0m_pc^2r_0\beta^2_0.
\label{eq:Munoz.3pmm}
\end{equation}
In SXF lattice description files (discussed further in the 
accompanying paper\cite{AGSAnalogue})
the RF parameter $V$ actually stands for $10^{-9}$
times the maximum energy change in eV of charge $e$ passing
through the RF cavity.

One shows easily that both the total energy
\begin{equation}
\mathcal{E} 
 =
\Big(\gamma^Im_pc^2 - \frac{k}{r}\Big) + \frac{k}{r_0} 
 \equiv
\mathcal{E}_M + \frac{k}{r_0}, 
\label{eq:Munoz.3qm}
\end{equation}
(where the ``I'' superscript on $\gamma^I$
re-emphasizes that this relativistic factor has to be evaluated 
``inside'' the bend element, where it depends on the local electric
potential) and the angular momentum
\begin{equation}
L = \gamma^Im_pr^2\dot\theta,
\label{eq:Munoz.3q}
\end{equation}
are constants of the motion. It has been necessary to
distinguish between the Mu\~noz potential energy $eV_M$ and our 
potential energy $eV$, because our potential vanishes on the
design orbit, while theirs vanishes at infinity. 
For the design orbit, using Eqs.~(\ref{eq:Munoz.3pmm})
and (\ref{eq:Munoz.3qm}), one obtains
\begin{equation}
\mathcal{E}_{M,0}
 =
\frac{m_pc^2}{\gamma_0}.
\label{eq:Munoz.3r}
\end{equation}
(The curious presence of $\gamma_0$ in the denominator
follows from the definition of potential energy.)
Using the obvious unit vector relations 
(valid for motion with angle $\theta$ increasing)
\begin{equation}
{\hat{\bm\theta}}'
 =
-{\bf\hat{r}},
\quad\hbox{and}\quad
{\bf\hat{r}}'
 =
{\hat{\bm\theta}},
\label{eq:Munoz.4}
\end{equation}
where primes stand for $d/d\theta$, the Lorentz equation can be
re-expressed as
\begin{equation}
\frac{d{\bf p}}{dt}
 =
\frac{k}{r^2}\,
\frac{d{\hat{\bm\theta}}}{d\theta}.
\label{eq:Munoz.5}
\end{equation}
Defining covariant velocity ${\bf u}=\gamma^I{\bf v}$,
and using Eq.~(\ref{eq:Munoz.3q}) to express $d\theta/dt$ 
in terms of $L$, the ${\bf u}$ equation of motion is
\begin{equation}
\frac{d{\bf u}}{d\theta}
 =
\frac{k\gamma^I}{L}\,
\frac{d{\hat{\bm\theta}}}{d\theta}.
\label{eq:Munoz.6}
\end{equation}
Mu\~noz and Pavic then introduce the generalized Hamilton
vector
\begin{equation}
\widetilde {\bf h} 
 =
{\bf u} - \frac{k\gamma^I}{L}\,{\hat{\bm\theta}},
\quad\hbox{or}\quad
(\widetilde h_{\theta},\widetilde h_r)
 =
\Big(u_{\theta}-\frac{k\gamma^I}{L}, u_r\Big),
\label{eq:Munoz.7}
\end{equation}
which we refer to as the MP-vector. It
is specially tailored so that the 
differential equation for $\widetilde {\bf h}$ reduces to
\begin{equation}
\frac{d\widetilde {\bf h}}{d\theta}
 =
-\frac{k}{L}\,\frac{d\gamma^I}{d\theta}\,
{\hat{\bm\theta}}.
\label{eq:Munoz.8}
\end{equation}
(The previously mentioned constancy of $\widetilde{\bf h}$ on
circular orbits is evident from this equation, since $\gamma^I$
is constant on, and only on, circular orbits.) Using
\begin{equation}
u_{\theta}
 = 
\frac{L}{m_pr},
\quad\hbox{}\quad
\widetilde h_{\theta}
 =
\frac{L}{m_pr}
 -
\frac{k\gamma^I}{L}
\overset{\rm also}{\ =\ }
\frac{\kappa^2L}{m_pr}
-\frac{k}{L}\,\frac{\mathcal{E}_M}{m_pc^2},
\label{eq:Munoz.8p}
\end{equation}
where the new parameter $\kappa$ (soon to be
identified as horizontal betatron ``tune'')
is defined by
\begin{equation}
\kappa^2
 =
1 - \bigg(\frac{k}{Lc}\bigg)^2.
\label{eq:Munoz.10}
\end{equation}
On the design orbit 
\begin{equation}
\kappa_0
 = 
\sqrt{1 - \beta_0^2}
 = 
\frac{1}{\gamma_0}
\quad 
( =\ 0.801213\ {\rm for\ pEDM}).
\label{eq:Munoz.10p}
\end{equation}
(Of course $\kappa_0$ will not be controlled to this accuracy
in practice and, in any case, quadrupoles and drifts in the lattice will
alter the ring tunes.) Solving Eq.~(\ref{eq:Munoz.8p}) for $r$,
the orbit equation in bend elements
can be expressed in terms of $\widetilde h_{\theta}$;
\begin{equation}
r 
 =
\frac{\lambda}{1+\bar\epsilon \widetilde h_{\theta}},
\label{eq:Munoz.8q}
\end{equation}
where
\begin{equation}
\lambda
 =
\frac{L^2c^2\kappa^2}{k\mathcal{E}_M}
\quad\hbox{and}\quad
\bar\epsilon
 =
\frac{m_pc^2}{\mathcal{E}_M}\,
\frac{L}{k}.
\label{eq:Munoz.8r}
\end{equation}
The parameter $\lambda$ is especially important in the storage ring
context, because the second term in the denominator of Eq.~(\ref{eq:Munoz.8q})
is both small compared to the first, and oscillatory. As a result $\lambda$
can be thought of loosely as an average radial position. 
(As earlier, the ``constants of motion'' $L$, $\kappa$, $\bar\epsilon$, 
and $\lambda$ are to be referred to as ``orbit elements''. Though
constant along any single particle orbit, they are different (though only
very slightly) for each individual particle.)

In component form Eq.~(\ref{eq:Munoz.8}) is
\begin{equation}
\widetilde h_r' - \widetilde h_{\theta}=0,
\quad
\widetilde h_{\theta}' + \widetilde h_r = -\frac{k}{L}\,{\gamma^I}'.
\label{eq:Munoz.9}
\end{equation}
The relativistic factor $\gamma^I$ 
(equal to $(\mathcal{E}_M+k/r)/(m_pc^2)$) can be expressed 
in terms of $\widetilde h_{\theta}$. First, eliminate $r$ in favor
of $u_0$ using the first of Eqs.~(\ref{eq:Munoz.8p}), then eliminate
$u_0$ if favor of $\widetilde h_{\theta}$, and, finally, solve
for $\gamma^I$;
\begin{equation}
\gamma^I
 =
\frac{\mathcal{E}_M}{\kappa^2m_pc^2}
 +
\frac{k}{\kappa^2Lc^2}\,\widetilde h_{\theta}.
\label{eq:Munoz.11}
\end{equation}
Needed for simplifying Eqs.~(\ref{eq:Munoz.9}), 
differentiating this equation yields
\begin{equation}
{\gamma^I}'
 =
\frac{k}{\kappa^2Lc^2}\,\widetilde h'_{\theta}.
\label{eq:Munoz.11p}
\end{equation}
(Heuristically, this expresses the dependence of mechanical energy
accompanying the change in potential energy associated with
change in particle position.)
After several lines of algebra, Eqs.~(\ref{eq:Munoz.9})
then reduce to
\begin{equation}
   \boxed{
   \begin{aligned}
      \widetilde h_r' &= \widetilde h_{\theta},  \label{eq:Munoz.12} \\
      \widetilde h_{\theta}' &= -\kappa^2\,\widetilde h_r.
   \end{aligned}
   }
\end{equation}
These are the equations that justify having introduced the
generalized Hamilton vector. Their general solution
can be written as
\begin{align}
\widetilde h_{\theta} 
 &=
\widetilde{\mathcal{C}}\,\cos\kappa(\theta-\theta_0) \notag\\
\widetilde h_r 
 &=
\frac{\widetilde{\mathcal{C}}}{\kappa}\,\sin\kappa(\theta-\theta_0).
\label{eq:Munoz.13}
\end{align}
where $\theta$ is a running angle in the interior of the
bend and $\theta_0$ is an angle to be determined, along with
$\widetilde{\mathcal{C}}$, by matching to the known initial conditions.
These new ``orbit elements'', $\mathcal{C}$ and $\theta_0$, 
are expressible in terms of the
previously defined orbit elements.
They are now to be used to parameterize
the true, particle-by-particle orbits through
the bend element.
(As explained elsewhere, the effects of the small deviations
of the actual electric field from the Coulomb field are to
be corrected for by virtual delta function kicks 
between bend slices.)
Concentrating on $\widetilde h_{\theta}$, and \emph{defining} $\theta$
to be zero at the bend entrance, we have
\begin{align}
\widetilde h_0 &\equiv \widetilde h_{\theta}|_{\theta=0} = \widetilde{\mathcal{C}}\cos\kappa\theta_0,\notag\\
\widetilde h'_0 &\equiv \widetilde h'_{\theta}|_{\theta=0} = \widetilde{\mathcal{C}}\kappa\sin\kappa\theta_0.
\label{eq:Munoz.13p}
\end{align}
These equations determine $\widetilde{\mathcal{C}}$ and $\theta_0$ to satisfy
\begin{equation}
\widetilde{\mathcal{C}} = \sqrt{{\widetilde h}_0^2 + \frac{({\widetilde h_0}')^2}{\kappa^2}},
\quad\hbox{and}\quad
\tan\kappa\theta_0 
 =
\frac{{{\widetilde h}'_0}}{\kappa \widetilde h_0}.
\label{eq:Munoz.13pp}
\end{equation}
The initial conditions for ${\widetilde h}_{\theta}$ need to be determined 
from the known coordinates of the particle just as it enters the
bend region. From its defining Eq.~(\ref{eq:Munoz.7}),
\begin{equation}
{\widetilde h}_0
 =
u_0 - \frac{k\gamma^I}{L},
\label{eq:Munoz.7m}
\end{equation}
Note that $\gamma$, like $L$, is
a running, particle-specific probe variable. updated on every 
entry to or exit from a bend element. 
When inside the bend it is equal to $\gamma^I$, but $\gamma$
varies discontinuously as the particle passes bend
edges (off-axis). In Eq.~(\ref{eq:Munoz.8p}), for any particular particle, 
$r$ is the only factor which is not a constant of the motion. 
We therefore have
\begin{equation}
{{\widetilde h}'}_0
 =
\frac{d\widetilde h}{d\theta}\Big|_{\theta=0}
 =
-\kappa^2\,\frac{L}{m_pr^2}\,\frac{dr}{d\theta}\Big|_{\theta=0}
 =
-\kappa^2\,\frac{L}{m_pr^2}\,r_0\frac{dx}{ds}\Big|_{\theta=0}.
\label{eq:Munoz.7p}
\end{equation}
The second of Eqs.~(\ref{eq:Munoz.13pp}) can then be used to fix 
$\theta_0$, but this requires resolving the multiple-valued 
nature of the inverse tangent function. 
One hopes this can be done by 
requiring $r$ to be ``smooth'' (continuous, with continuous slope) 
across the bend entrance edge. Even this is tricky since, for 
non-normal incidence, there is a refractive 
deflection\footnote{A ``refractive deflection'' is mentioned here
and at other places in the paper. With ``hard edge'' bending 
elements the electric potential changes discontinuously from
just outside to just inside the edges of the element. The mechanical
energy has a corresponding discontinuity and, as a result, so also
does the magnitude of the momentum (which is predominantly 
longitudinal). But, since there is no transverse force, the
transverse momentum is continuous across the boundary. Taken
together, the result is an angular (Snell's law like) refractive
angular deflection of each orbit at each field edge. Similar
deflections occur at quadrupole edges. But, as well as being
far weaker, the entry and exit deflections tend to cancel in 
all thin elements. In ETEAPOT
the refractive corrections are included at all bend edges and
neglected at all other edges. At bend field edges there are
also fringe field corrections (which are especially important
sources of spin decoherence.) The fringe field and refractive
corrections are evaluated separately and simply superimposed
at every bend edge.}
at the boundary. This deflection is small, especially near normal 
incidence, which is always true in our case.
To resolve the inverse tangent ambiguity for non-normal incidence
it would usually be sufficient to choose the value for which the 
slope is most nearly continuous. But to be safe one has to 
account explicitly for the refractive kink. 

From Eq.~(\ref{eq:Munoz.8q}) the radial coordinate $r$ is
given by
\begin{equation}
r = \frac{\lambda}{1+\epsilon\cos\kappa(\theta-\theta_0)},
\quad\hbox{where}\quad
\epsilon=\widetilde{\mathcal{C}}\bar\epsilon.
\label{eq:Munoz.14}
\end{equation}
This is the orbit equation previously introduced in Eq.~(\ref{eq:Munoz.14m}). 
Borrowing terminology from solar planetary orbits about the sun, from its 
structure, one sees that $\theta_0$ is either the angle at perihelion, 
or aphelion, depending on the sign of $\epsilon$. 
In multi-million turn tracking, there are
many opportunities to misidentify aphelion and 
perihelion or to obtain the wrong sign for $\epsilon$---this
complication is associated with the degeneracy of the central orbit 
that was mentioned earlier.

Using Eqs.~(\ref{eq:Munoz.8r}), 
(\ref{eq:Munoz.13pp}), and (\ref{eq:Munoz.14}),
\begin{equation}
\epsilon
 =
\widetilde{\mathcal{C}}\,
\frac{m_pc^2}{\mathcal{E}_M}\,
\frac{L}{k}
 =
\sqrt{{\widetilde h}_0^2 + \frac{({\widetilde h_0}')^2}{\kappa^2}}\,
\frac{m_pc^2}{\mathcal{E}_M}\,
\frac{L}{k}.
\label{eq:Munoz.14p}
\end{equation}
By expressing the small parameter $\epsilon$, proportional to the
small Hamilton vector value and slope (or rather their quadrature sum), 
this formula is essential to the implementation of the Mu\~noz/Pavic approach.
(These ``small'' quantities can even be infinitesimal in the 
analytic treatment.) 
 
From these equations one has obtained simple analytic 
parameterizations for $x(\theta)$ and $\gamma^I(\theta)$.
In particle tracking one knows $x_{\rm in}$ to high accuracy
and wishes to find $x_{\rm out}$, also to high accuracy. 
Their difference is given by
\begin{align}
x_{\rm out} - x_{\rm in}
 &=
\frac{\lambda}{1+\bar\epsilon\,{\widetilde h}_{\theta}} - r_0
 -
\frac{\lambda}{1+\bar\epsilon\,{\widetilde h}_0} + r_0 \notag\\
 &=
\frac{{\widetilde h}_0 - {\widetilde h}_{\theta}}
     { (1+\bar\epsilon\,{\widetilde h}_{\theta})(1+\bar\epsilon\,\widetilde h_0)}\,
\lambda\bar\epsilon.
\label{eq:Munoz.14pp}
\end{align}
This manipulation has suppressed the harmful cancellation exhibited
in Eq.~(\ref{eq:Munoz.14}) which gives $r$ rather that the radial
displacement from the design orbit. 
Eventually, when deviation from $1/r^2$ electric field dependence 
is modeled by thin virtual quadrupoles, to improve precision, one
will slice the bends finer and finer. 

In the context of accelerator physics one can ask for the
similar evolution equations for vertical coordinate $y$. But
this question is inappropriate; the Kepler orbit lies, by
definition, in a single plane and we always choose the
propagation plane to be the plane containing the incident velocity
vector. Instead of keeping track of $y$ we have to keep track
of the orbit plane or, equivalently, the normal to the orbit
plane. This vector changes discontinuously (but only by
a tiny angle) in passing through thin multipole elements. It 
is undefined in drift regions and is constant in bend elements.
Keeping track of the orbit plane is covered in later sections.

\paragraph{\bf Rescaling of the MP-vector and updating the horizontal slope.}
For updating the horizontal slope component,
we introduce the previously-announced rescaling of the 
MP-vector;
\begin{equation}
{\bf h} \equiv \frac{\widetilde{\bf h}}{\gamma_0v_0}.
\label{eq:BetaFuncts.1}
\end{equation}
Dividing by the design velocity $v_0$ has rendered ${\bf h}$
dimensionless, and the further factor of $\gamma_0$ in the
denominator simplifies matching to conventional lattice function
representation.  At this time
we also switch definitions of the prime operator notation
so that $f'\equiv df/ds$, where
$s$ is arc length along the design orbit. With this revised
notation Eqs.~(\ref{eq:Munoz.12}) become
\begin{equation}
   \boxed{
   \begin{aligned}
      h_r' &= \frac{1}{r_0}\, h_{\theta},  \label{eq:Munoz.12pp} \\
      h_{\theta}' &= -\frac{1}{r_0}\,\kappa^2\, h_r.
   \end{aligned}
   }
\end{equation}
Finally it is possible to make contact with more familiar
accelerator variables, such as $p_x$ and $p_s$, the radial and
longitudinal particle momentum components.

From the definition of $\tilde h_r$ in Eq.~(\ref{eq:Munoz.7}),
after re-scaling~(\ref{eq:BetaFuncts.1}), the transverse momentum
is given by
\begin{equation}
p_x = \gamma_0m_pv_0\,h_r,
\quad\hbox{or}\quad
x' \equiv \frac{dx}{ds} = {\tt p[1]} = \frac{p_x}{p_z} = h_r,
\label{eq:BetaFuncts.1p}
\end{equation}
where $p_0=\gamma_0m_pv_0$ is the design momentum and 
{\tt p[1]} is the symbol used for $x'$ internally in the UAL code.
(Some components in the chain of equalities in 
Eq.~(\ref{eq:BetaFuncts.1p}) may assume paraxial approximation, even
if for no reason other than that the definition of $x$ is unambiguous
only in the paraxial limit.)
So $h_r$ is nothing other than the phase space coordinate conjugate
to $x$, namely $dx/ds$. 

When applied at the exit of a bend,
the value of $x'$ given by Eq.~(\ref{eq:BetaFuncts.1p})
would better have been called $x'_{\rm out,-}$, since the refractive 
compensation associated with exiting the bend would still need to be made to 
produce {\tt p[1]} valid just outside the bend exit. 

Potential loss of numerical precision is always an issue. 
A compact, numerically accurate way to update the 2D phase space 
coordinates is to work
directly from Eq.~(\ref{eq:Munoz.14}). For bend angle $\Delta\theta$,
the argument $\theta-\theta_0$ is $\Delta\theta$. The increment in $x$
from input to output is
\begin{equation}
x_{\rm out} - x_{\rm in}
 = 
\lambda\epsilon\,\frac{\cos\kappa\theta_0 - \cos\kappa\Delta\theta}
     {(1+\epsilon\cos\kappa\Delta\theta)(1+\epsilon\cos\kappa\theta_0)}
\label{eq:Compact.1}
\end{equation}
Differentiating Eq.~(\ref{eq:Munoz.14}) produces
\begin{equation}
\frac{dx}{d\theta}(\theta)
 = 
\lambda\epsilon\kappa\,\frac{\sin\kappa(\theta-\theta_0)}
     {\big(1+\epsilon\cos\kappa(\theta-\theta_0)\big)^2}.
\label{eq:Compact.2}
\end{equation}
Including the change of independent variable from $\theta$
to $z$, the increment from input to output is
\begin{align}
&\frac{dx}{dz}\Big|_{\rm out} - \frac{dx}{dz}\Big|_{\rm in} = \notag\\
  &
 = \frac{\lambda\epsilon\kappa}{r_0}\,
\bigg(
\frac{\sin\kappa\Delta\theta}
     {(1+\epsilon\cos\kappa\Delta\theta)^2}
 +
\frac{\sin\kappa\theta_0}
     {(1+\epsilon\cos\kappa\theta_0))^2}
\bigg)
.
\label{eq:Compact.3}
\end{align}
These formulas avoid taking the difference of nearly equal terms.

\paragraph{\bf Pseudoharmonic description of the motion.}
In general the MP plane will be close to, but not exactly identical
with the horizontal lattice design plane. In this section,
for simplicity in making contact with Twiss functions, these 
two planes will be treated as equivalent. 

At this point we wish to correlate the dynamic quantities
introduced so far with more familiar (to accelerator
scientists) quantities such as Twiss functions, betatron
phase advances, transfer maps, and so on. The relationships
are especially simple for a combined function, weak-focusing
ring; this will be described before developing the full
Twiss function formulation needed for separated function 
lattices. In general,
the definition of lattice functions within electric bend 
elements will be more complicated, but also not very
important for the EDM experiment.

Eq.~(\ref{eq:Munoz.12pp}) yields 
\begin{equation}
h_{\theta} = r_0\,\frac{d^2x}{ds^2}.
\label{eq:eq:BetaFuncts.1q}
\end{equation}
In traditional Courant-Snyder (CS) formalism the betatron phase $\psi$ 
and the positive-definite lattice function $\beta(s)$ are related by
\begin{equation}
\psi'
\equiv
\frac{d\psi}{ds}
 =
\frac{1}{\beta(s)},
\label{eq:BetaFuncts.2}
\end{equation}
Even with $\theta = s/r_0$, because $\beta(s)$ is not necessarily 
constant, the angles $\psi$ and $\theta=s/r_0$, though monotonically
related, are not strictly proportional.
The virtue of the beta function formalism is primarily due to
its ability to describe each orbits in a complicated lattice as
a sinusoidal function of unambiguous phase advance $\psi$.

Shortly we will have to admit that our generalization of the
CS formalism to electric lattices will be limited, in principle,
to describing orbits \emph{outside} bending elements, where the
electric potential vanishes. The parameter 
$\beta$ introduced now will be applied \emph{inside} bends, 
and it is assumed to be independent of $s$. Furthermore it will 
not match up smoothly with the adjacent \emph{outside} $\beta$ 
functions. This reflects the small discontinuity in particle 
kinetic energy at bend edges.

In practice this issue is largely academic. In any
realistic ring there are many drift spaces, more or less 
uniformly distributed. Especially with the weak focusing expected
in EDM rings, the ranges of $\beta$ function variations will be
quite limited. Simply interpolating the (slightly ambiguous)
beta functions in the bend regions from their (reliably known)
values in the drifts should be sufficient for most purposes. 

As already stated, our electric lattice model admits only 
uniform, inverse square law bending elements (though with artificial
thin trim elements). Within such a bend we approximate
$\beta'\equiv d\beta/ds=0$ 
and horizontal``betatron'' oscillations of amplitude $a$ are 
described by
\begin{align}
x  &=  a\beta^{1/2}\cos\psi, \notag \\
x' &= -a \beta^{1/2}\sin\psi\,\psi'
    = -a \beta^{-1/2}\sin\psi.
\label{eq:BetaFuncts.3}
\end{align}
Differentiating once more yields
\begin{equation}
{x'}' = -a \beta^{-3/2}\cos\psi = -\beta^{-2}x.
\label{eq:BetaFuncts.4}
\end{equation}
Substituting this into Eq.~(\ref{eq:Munoz.12pp}) produces
\begin{equation}
x = -\frac{\beta^2}{r_0}\,h_{\theta}.
\label{eq:BetaFuncts.5}
\end{equation}
Just as $h_r$ and $h_{\theta}$ vary ``in quadrature'', so
also do $x$ and $x'$. The ratios of their maxima are
\begin{equation}
\frac{h_{\theta,{\rm max}}}{h'_{\theta,{\rm max}}}
 =
\frac{r_0}{\kappa},
\quad\hbox{and}\quad
\frac{x_{\rm max}}{x_{\rm max}'}
 =
\frac{h_{\theta,\rm max}}{h_{\theta,\rm max}'}
 =
\frac{r_0}{\kappa} 
\overset{\rm also}{=}
\beta.
\label{eq:BetaFuncts.6}
\end{equation}
Collecting results, we have
\begin{equation}
\beta = \frac{r_0}{\kappa},
\quad
x = -\frac{r_0}{\kappa^2}\,h_{\theta},
\quad\hbox{and}\quad
x' = h_r.
\label{eq:BetaFuncts.7}
\end{equation}
(Remember that these $x$ and $x'$ values are
only the precise Frenet-Serret laboratory coordinates when the M-P plane 
coincides with the horizontal laboratory reference frame, which is
always a good approximation.)
To summarize, one sees that, to linearized approximation, the 
components of the MP-vector are, except for scale factors, identical to 
the betatron phase space components. 

The parameter $\beta$ has been used in this section only to
produce familiar-looking formulas. For a weak focusing ring we now see
that this $\beta$ need never have been introduced, since it is just an abbreviation for 
$r_0/\kappa$. It does not depend explicitly on $s$, 
but through its $\kappa$ factor
it depends on $\gamma$, $x$ and $x'$, and is therefore different for different
particles.
 
\paragraph{\bf Revolution period.}
For modeling longitudinal dynamics it is critical to obtain the
revolution period $T_{\rm rev.}$ to good accuracy for every particle.  
From Eqs.~(\ref{eq:Munoz.3q}), (\ref{eq:Munoz.11}), (\ref{eq:Munoz.13}),
and (\ref{eq:Munoz.14}) one has
\begin{align}
\frac{dt}{d\theta}
  &=
\frac{\mathcal{E}_M\lambda^2}{L\kappa^2c^2}\,
\frac{1}{\Big(1+\epsilon\cos\kappa(\theta-\theta_0)\Big)^2}
 + \notag\\
 &+ \frac{m_pk\widetilde{\mathcal{C}}\lambda^2}{\kappa^2L^2c^2}\,
\frac{\cos\kappa(\theta-\theta_0)}
{\Big(1+\epsilon\cos\kappa(\theta-\theta_0)\Big)^2}.
\label{eq:RevPeriod.1}
\end{align}
To obtain time of flight $dt/d\theta$ has to be integrated over 
$\theta$. Both integrals can be evaluated in closed form to obtain 
$t$ as a function of $\theta$. However the formulas are complicated
and their evaluation is numerically treacherous, for reasons
discussed below. 

For the study of longitudinal dynamics and synchrotron 
oscillations what is needed is the deviation of the time of flight 
from the time of flight of the design particle. This is a very small 
difference of two large numbers. Eq.~(\ref{eq:RevPeriod.1}) can be 
recast as
\begin{align}
t - t_0
 &=
A 
\int
\frac{d\Delta\theta}{(1+\epsilon\cos\kappa\Delta\theta)^2}
 - A_0
\int d\Delta\theta\,
 +   \notag\\
 &\qquad +
B\tilde{\mathcal{C}}
\int 
\frac{\cos\kappa\Delta\theta\, d\Delta\theta\,}
{(1+\epsilon\cos\kappa\Delta\theta)^2}.
\label{eq:RevPeriod.2}
\end{align}
Here $\Delta\theta=\theta-\theta_0$, the factors $A$ and $B$
(both of which deviate little from constancy and can never 
become ``small'')  
replace the other factors in Eq.~(\ref{eq:RevPeriod.1}),
and $A_0$ is the values of $A$ on the
design orbit. All of the factors $A$, $B$, $A_0$, $\epsilon$,
and $\kappa$ are constant on any particular orbit being followed, 
and all have values very close to their values on the design
orbit. The only variable factor is $\Delta\theta$.
The first two terms cancel approximately, which
makes the evaluation of the first term critical. The third
term requires no special treatment since the $\tilde{\mathcal{C}}$
factor is already differentially small. The cancellation tendency 
can be ameliorated by adding and subtracting a term 
$A_0\int d\Delta\theta=A_0\Delta\theta$ to produce
\begin{align}
t - t_0
 &=
\epsilon
\int
\frac{(-2A + B/\tilde{\epsilon})\cos\kappa\Delta\theta
      -A\epsilon\cos^2\kappa\Delta\theta}
{(1+\epsilon\cos\kappa\Delta\theta)^2}\,
d\Delta\theta +   \notag\\
 &\qquad + 
(A-A_0)\Delta\theta.
\label{eq:RevPeriod.3}
\end{align}
Here $\tilde{\mathcal{C}}=\epsilon/\tilde\epsilon$ has been 
obtained using Eq.~(\ref{eq:Munoz.14}). 
In Eq.~(\ref{eq:RevPeriod.3}) all terms are differentially 
small and the small differences of large numbers have been moved outside
the integrals. For the proton EDM experiment, since $|\epsilon|$
never exceeds $1/1000$, the integrand can be expanded in powers 
of $\epsilon$ before integrating.  A typical leading term in these
integrals is $\Delta\theta$ multiplied by a constant factor. 
While using this time of flight formalism  
ETEAPOT includes this term as well as terms with
extra powers of $\epsilon$ up to $\epsilon^5$ and \emph{does not} check 
automatically that this last term is, itself, negligible.
An alternate, seemingly more robust, hybrid time of flight calculation,
now used by default in ETEAPOT, is desribed later. 

A subtle feature of the time of flight evaluation results from
the fact that 
the ellipticity $\epsilon$ vanishes on perfectly circular orbits
(of which the design orbit is one). This causes the angle $\theta_0$,
which is the angle to perigee, to be indeterminate because
perigee is indeterminate on a circular orbit. Eq.~(\ref{eq:Munoz.14}),
the second of Eqs.~(\ref{eq:Munoz.8r}) and 
Eq.~(\ref{eq:Munoz.13pp}) force $\epsilon$ to be non-negative.
And yet the major and minor axes can switch from prolate to
oblate when an arbitrarily small kink is applied to an orbit 
for which $\epsilon$ is close enough to zero. When this happens
the angle $\theta_0$ advances discontinuously through $\pi/2$. When
this happens the angle $\theta$ also advances discontinuously
through the same angle. 

These changes do not affect the
integrals in Eq.~(\ref{eq:RevPeriod.3}) which depend only
on $\theta-\theta_0$. They do, however, make the time of flight
calculation hyper-sensitive to the determination of $\theta_0$,
which can change erratically in progressing from one bend 
element to the next, after passing through a straight section
which possibly contains a quadrupole. A discontinuous change
in $\theta_0$ can change the analytic, closed-form indefinite
integrals used in applying Eq.~(\ref{eq:RevPeriod.1}). 

\section{Transformation from local frame to MP frame.}
\paragraph{\bf Determination of the MP bend plane.}
In order to reserve $x$ for radial displacement in the 
Mu\~noz-Pavic (MP) bend
plane we now use overhead bars to designate conventional
Courant-Snyder (CS) coordinates. At a location where the longitudinal
coordinate has been shifted to be $z=0$, the particle position 
$(\bar x, \bar y, 0)$ is given by
\begin{equation}
\frac{\bf r}{r_0}
 =
\Big(1 + \frac{\bar x}{r_0}\Big)\,{\bf\hat{\bar x}}
 + 
\frac{\bar y}{r_0}\,{\bf\hat{\bar y}}.
\label{eq:CStoMP.1}
\end{equation}
Scaled momentum components 
$\tilde p_{\bar x}=(p_{\bar x}/p_0)/(1+\delta p/p_0)$,
and
$\tilde p_{\bar y}=(p_{\bar y}/p_0)/(1+\delta p/p_0)$,
are defined by
\begin{equation}
\tilde{\bf p}
 =
\tilde p_{\bar x}{\bf\hat{\bar x}}
 +
\tilde p_{\bar y}{\bf\hat{\bar y}}
 +
\Big( 
1 - \frac{\tilde p^2_{\bar x}}{2}
 - \frac{\tilde p^2_{\bar y}}{2}
\Big)\,{\bf\hat{\bar z}},
\label{eq:CStoMP.2}
\end{equation}
where terms of fourth order and beyond have been dropped.
(For the EDM experiment such terms have numerical values
such as $(0.01/20)^4\approx10^{-13}$.) This has been arranged
so that $\tilde{\bf p}$ is a unit vector. The
${\tt p[1]}$ and ${\tt p[3]}$ MAD/UAL coordinates being 
tracked in the computer are close to, but not exactly
equal to $\tilde p_{\bar x}$ and $\tilde p_{\bar y}$.
In fact 
\begin{equation}
\tilde p_{\bar x} = \frac{{\tt p[1]}}{1+\delta p/p_0},
\quad\and\quad
\tilde p_{\bar y} = \frac{{\tt p[3]}}{1+\delta p/p_0}.
\label{eq:CStoMP.2p}
\end{equation}
Scaled angular momentum components are then given by
\begin{align}
& {\bf\tilde L}
 =
\frac{\bf L}{r_0p_0}
  =
\frac{\bf r\times{\bf p}}{r_0p_0}
 =
\det
\begin{pmatrix}
{\bf\hat{\bar x}} & {\bf\hat{\bar y}} & {\bf\hat{\bar z}} \\
1 + \bar x/r_0    &    \bar y/r_0     &     0             \\
\tilde p_{\bar x}  & \tilde p_{\bar y}  & \tilde p_{\bar z}
\end{pmatrix}                                      \notag\\
 &=
\frac{\bar y}{r_0}\,
\tilde p_{\bar z}
\,{\bf\hat{\bar x}}
 -
\Big(
1+\frac{\bar x}{r_0}
\Big)\,
\tilde p_{\bar z}\,
{\bf\hat{\bar y}}
 +
\Big(
\tilde p_{\bar y}
 + \frac{\bar x}{r_0}\,\tilde p_{\bar y}
 - \frac{\bar y}{r_0}\,\tilde p_{\bar x}
\Big)\,
{\bf\hat{\bar z}}.
\label{eq:CStoMP.3}
\end{align}
The scaled central angular momentum is 
$
{\bf\tilde L_0}
 =
-{\bf\hat{\bar y}}
$,
and the perpendicular component ${\bf\tilde L}_{\perp}$ is
\begin{align}
{\bf\tilde L}_{\perp}
 &=
\frac{\bar y}{r_0}\,
\Big( 
1 - \frac{\tilde p^2_{\bar x}}{2}
 - \frac{\tilde p^2_{\bar y}}{2}
\Big)
\,{\bf\hat{\bar x}}
 +
\Big(
\tilde p_{\bar y}
 + \frac{\bar x}{r_0}\,\tilde p_{\bar y}
 - \frac{\bar y}{r_0}\,\tilde p_{\bar x}
\Big)\,
{\bf\hat{\bar z}} \notag \\
 &\equiv
\sin\phi_{\bar z}
\,{\bf\hat{\bar x}}
 -
\sin\phi_{\bar x}
{\bf\hat{\bar z}},
\label{eq:CStoMP.5}
\end{align}
which defines two (small) angles the tilted bend plane
makes with true vertical. This equation determines the
(small) rotation angles about transverse axes that rotate
the horizontal plane into the MP plane. The trigonometry
is not exact. But it is accurate enough
for the extremely small angles involved.

\paragraph{\bf Merging 2D MP-tracking into 3D.}
With the orbit geometry being worked out in the 2D MP plane,
it remains necessary to transform back and forth to the 
local 3D Frenet frame. The goal of a tracking calculation is to 
obtain the output particle radius vector ${\bf r}_{\rm out}$
for a given input particle radius vector ${\bf r}_{\rm in}$.  The 
corresponding unit vector evolution is from 
${\bf\hat r}_{\rm in}=(\hat x_{\rm in},\hat y_{\rm in},0)$ to 
${\bf\hat r}_{\rm out}$. We calculate this unit vector evolution
first.

On the design orbit the input and output
vectors are ${\bf\hat d}_{\rm in}=(1,0,0)$ and 
${\bf\hat d}_{\rm out}=(\cos\theta,0,\sin\theta)$, both lying
in the horizontal $(xz)$-plane. With the design bend angle
defined to be $\theta$, one has
\begin{equation}
{\bf\hat d}_{\rm in}\cdot{\bf\hat d}_{\rm out}
 =
\cos\theta.
\label{eq:Fully3D.1}
\end{equation}
The particle-specific bend angle $\phi$, in its own
bend plane, satisfies
\begin{equation}
{\bf\hat r}_{\rm in}\cdot{\bf\hat r}_{\rm out}
 =
\cos\phi,
\label{eq:Fully3D.2}
\end{equation}
where $\phi$ is close to, but not equal to, $\theta$ in general.
Letting $a$ and 
$b=\pm|b|=\pm\sqrt{1-a^2}$ be the components of ${\bf\hat r}_{\rm out}$
along ${\bf\hat d}_{\rm out}$, one has
\begin{equation}
{\bf\hat r}_{\rm out}
 =
(a\,\cos\theta, \pm\sqrt{1-a^2}, a\sin\theta).
\label{eq:Fully3D.3}
\end{equation}
(For clockwise orbit) the unit vector normal to the 
bend plane,
\begin{equation}
{\bf\hat n}
 =
-\frac{{\bf L}_{\rm in}}{|{\bf L}_{\rm in}|},
\label{eq:Fully3D.4}
\end{equation}
is known from calculations based on the particle coordinates
at the entrance to the bend element. 
To determine $a$ we require 
\begin{equation}
0
 =
{\bf\hat n}_{\rm in}\cdot{\bf\hat r}_{\rm out}
 =
n_xa\,\cos\theta \pm n_y\sqrt{1-a^2} + n_za\,\sin\theta.
\label{eq:Fully3D.5}
\end{equation}
Solving for $a$,
\begin{equation}
a
 =
\frac{n_y}
{\sqrt{n_y^2 + n_x^2\cos^2\theta + 2 n_xn_z\cos\theta\sin\theta + n_z^2\sin^2\theta}}.
\label{eq:Fully3D.6}
\end{equation}
The angular advance $\phi$ then satisfies
\begin{equation}
\cos\phi
 =
{\bf\hat r}_{\rm in}\cdot {\bf\hat r}_{\rm out}
 =
\hat x_{\rm in} a\cos\theta + \hat y_{\rm in} b.
\end{equation}
The radial coordinate is then given by Eq.~(\ref{eq:Munoz.8q})
which is repeated here;
\begin{equation}
r 
 =
\frac{\lambda}{1+\bar\epsilon \widetilde h_{\theta}},
\label{eq:Fully3D.7m}
\end{equation}
where $\widetilde h_{\theta}$ is given by Eq.~(\ref{eq:Munoz.13}).
The output radius vector is then given by
\begin{equation}
{\bf r}_{\rm out}
 =
r\,{\bf\hat r}_{\rm out}.
\label{eq:Fully3D.7}
\end{equation}
In this way the evolution can be completed using just
Courant Snyder coordinates.

\section{Transformed dependent variable $\xi$ orbit representation}
\paragraph{\bf Electric sector bends with field index $m$.}
The Mu\~noz and Pavic, Hamilton vector approach
described so far, though ideal for fast exact particle tracking, 
is less well matched to developing the Courant-Snyder
Twiss function description conventional in accelerator theory. 
A more conventional approach is taken in this section. As well
as enabling a truncated power series formalism for electric
rings, the results can be used for corroborating results already
obtained. This includes independently confirming the
integrable dynamics demonstrated for $m$=1. Though discovered using
the Mu\~noz/Pavic approach, this remarkable integrability 
property must necessarily
be derivable also in the paraxial-based approach. Initially this 
treatment will be generalized to include field indices other than
$m=1$.

To produce weak vertical focusing the radial electric field 
has to fall off as $1/r^{1+m}$, with $m>0$.
The $m=0$ case is singular, and leads to a logarithmic 
potential.\footnote{Integrating 
the electric field from $r=1$ to $r=1+\Delta$ one reconstructs 
the potential from the electric field---call it $\tilde{V}$. 
Its value at  $r=1+\Delta$ is
\begin{equation}
\tilde V(1+\Delta)
 = 
\frac{(1+\Delta)^{- m }-1}{ m }
 =
\ln(1-\Delta) + O( m ).
\label{eq:WeakFoc.0}
\end{equation}
The logarithmic potential can be seen to be a 
degenerate form required for $m=0$, where $E\propto1/r$. The electric 
field corresponding to this potential is $E\propto1/r^{1+ m }$. 
This shows how $\tilde{V}$ approaches the logarithmic potential
in the limit of small $ m $.
}
Introducing design radius $r_0$ and central field $-E_0$
the electric field for $y$=0 is
\begin{equation}
{\bf E}(r,0)
 = 
-E_0\,\frac{r_0^{1+ m }}{r^{1+ m }}\,{\bf\hat r}.
\label{eq:WeakFoc.1}
\end{equation}
The electric potential $V(r)$, adjusted to vanish 
at $r=r_0$ was given earlier in Eq.~(\ref{eq:WeakFoc.2m}).
The independent (longitudinal) coordinate $s$ is to be 
replaced by the angular coordinate $\theta$
\begin{equation}
\theta = \frac{s}{r_0}.
\label{eq:WeakFoc.2pmm}
\end{equation}
Following standard treatments of relativistic Kepler 
orbits\cite{Moller-SR}\cite{Aguirregabiria}\cite{Torkelsson}\cite{Boyer}, 
we change dependent variable from $x(s)=r-r_0$ 
(with independent variable $s$) to 
a (dimensionless) dependent variable $\xi(\theta)$ (with  
independent variable $\theta$);
\begin{align}
\xi &=  \frac{x}{r} = \frac{x}{r_0+x} = 1 - \frac{r_0}{r}, \notag\\
\frac{d\xi}{d\theta}
 &= 
\frac{r_0}{r^2}\,\frac{dr}{d\theta},
\quad\hbox{or}\quad
\xi'
 = 
\frac{r^2_0}{r^2}\,x'.
\label{eq:WeakFoc.2p}
\end{align}
Note that $\xi$ is proportional to $x$ for small $x$ and
that, notationally, $x'\equiv dx/ds$ and $\xi'=d\xi/d\theta$.
The present discussion is limited to planar orbits, in which case
the definition of $x$ by $r=r_0+x$ is the usual Frenet-Serret 
definition. For 3D motion this definition is not quite exact but
realistic vertical amplitudes will always be small enough that
the effect on $x$ of projection onto the horizontal plane will
be neglected. The identity 
\begin{equation}
\frac{r_0}{r} = 1-\xi
\label{eq:WeakFoc.2pp}
\end{equation}
will be prominent in subsequent formulas. 
Inverse relations are
\begin{equation}
x = \frac{r_0\xi}{1-\xi},
\quad\hbox{and}\quad
x'
 =
\frac{r_0 d\xi/ds}{(1-\xi)^2}
 =
\frac{\xi'}{(1-\xi)^2}.
\label{eq:WeakFoc.2q}
\end{equation}
With $\xi$ regarded as a function of $\theta$ and $x$ as 
a function of $s$, the abbreviations $\xi'=d\xi/d\theta$
and $x'=dx/ds$ have been used.

For simplicity in describing the approach to be taken,
we temporarily specialize to the $m$=0 ``cylindrical''
case. The electric potential $V(r)$, adjusted to vanish on the
design orbit, is 
\begin{equation}
V(r)
 = 
E_0r_0\ln\frac{r}{r_0} 
 =
  E_0r_0\ln r 
- E_0r_0\ln r_0.
\label{eq:Cylinder.2}
\end{equation}
The design parameters are related by
\begin{equation}
eE_0r_0
 = 
\beta_0p_0c
 =
m_pc^2
\Big(
\gamma_0 - \frac{1}{\gamma_0}
\Big),
\label{eq:Cylinder.3}
\end{equation}
where $p$, $v$, and $\beta$ are proton momentum, velocity,
and $v/c$. The momentum vector components are defined by
\begin{align}
{\bf p}
 &=
p_r{\bf\hat r} + p_{\theta}{\bf\hat\theta} + p_y{\bf\hat y}
\label{eq:Cylinder.3p}\\
 &=
 \frac{m_p\dot r}{\sqrt{1-v^2/c^2}}\,{\bf\hat r}
+\frac{m_pr\dot\theta}{\sqrt{1-v^2/c^2}}\,{\bf\hat\theta}
+\frac{m_p\dot y}{\sqrt{1-v^2/c^2}}\,{\bf\hat y}. \notag
\end{align}
The electric force alters only the radial momentum component
\begin{equation}
\frac{dp_r}{dt}
 =
-eE_0\frac{r_0}{r}.
\label{eq:Cylinder.3q}
\end{equation}
For $m$=0 the invariance of the system
to translation along the $y$-axis, causes $p_y$ to be
conserved, and to invariance under rotation around the central 
axis, causes $L_y$, the vertical component of angular momentum, 
to be conserved. For a particle in the horizontal plane
containing the origin the angular momentum vector is 
\begin{equation}
{\bf L}
 =
{\bf r}\times{\bf p}
 =
-\frac{m_pr\dot y}{\sqrt{1-v^2/c^2}}\,{\bf\hat\theta}
+\frac{m_pr^2\dot\theta}{\sqrt{1-v^2/c^2}}\,{\bf\hat y}
 =
L_{\theta}{\bf\hat\theta} + L_y{\bf\hat y}.
\label{eq:Cylinder.4}
\end{equation}
The design orbit angular momentum is 
\begin{equation}
L_0
 =
m_pc\beta_0\gamma_0r_0.
\label{eq:Cylinder.4pm}
\end{equation}
We seek the orbit differential equation giving dependent variable 
$r$ as a function of independent variable $\theta$.
The (conserved) total proton energy $\mathcal{E}$ 
is the sum of the mechanical energy and the potential energy
\begin{equation}
\mathcal{E}
 =
\sqrt{p_r^2c^2 + p_{\theta}^2c^2 + p_y^2c^2+m_p^2c^4} + eV(r).
\label{eq:Cylinder.5}
\end{equation}
Squaring this equation yields 
\begin{equation}
\big(\mathcal{E}-eV(r)\big)^2
 =
p_r^2c^2 + p_{\theta}^2c^2 + p_y^2c^2+m_p^2c^4.
\label{eq:Cylinder.6}
\end{equation}
The terms on the right hand side of this equation can be expressed 
in terms of $r$, $dr/d\theta$, and conserved quantities.
From Eqs.~(\ref{eq:Cylinder.3p}) and (\ref{eq:Cylinder.4}),
\begin{equation}
p_{\theta}
 =
\frac{L_y}{r}.
\label{eq:Cylinder.4p}
\end{equation}
$p_r$ can be expressed similarly
using Eq.~(\ref{eq:Cylinder.3p});
\begin{equation}
\frac{p_r}{p_{\theta}}
 =
\frac{\dot r}{r\dot\theta}
 =
\frac{1}{r}\,\frac{dr}{d\theta},
\quad\hbox{and\ hence}\quad
p_r
 =
\frac{L_y}{r^2}\,\frac{dr}{d\theta}.
\label{eq:Cylinder.7}
\end{equation}
Making these substitutions yields
\begin{equation}
\big(\mathcal{E}-eV(r)\big)^2
 =
\bigg(\frac{L_yc}{r^2}\frac{dr}{d\theta}\bigg)^2
 + \frac{L_y^2c^2}{r^2}
 + p_y^2c^2
 + m_p^2c^4.
\label{eq:Cylinder.8}
\end{equation}
\begin{widetext}
When the field index $m$ was first introduced it was visualized
as being a small number, just large enough (and positive) to produce
some vertical focusing. But the subsequent analysis has not relied on
$m$ being small.
Generalizing the previous formulas for arbitrary $m$ values, and
expressing the in-plane momentum in terms of angular momentum $L_y$,
$\xi$ and $\xi'$,
the mechanical (total minus potential) energy can be expressed in
terms of particle mass and momentum components. This generalizes 
Eq.~(\ref{eq:Cylinder.8}) for arbitrary $m$;
\begin{equation}
\bigg(
\mathcal{E} - \frac{eE_0r_0}{ m } +\frac{eE_0r_0}{ m }\,(1-\xi)^{ m }
\bigg)^2
 =
\frac{L_y^2c^2}{r_0^2}\bigg(\frac{d\xi}{d\theta}\bigg)^2
 + 
\frac{L_y^2c^2}{r_0^2}\,(1-\xi)^2 + p_y^2c^2 + m_p^2c^4.
\label{eq:WeakFoc.3}
\end{equation}
Differentiating this equation with respect to $\theta$, and
simplifying,
\begin{align}
&\frac{d^2\xi}{d\theta^2}
 =
 1-\xi 
-\bigg(
\mathcal{E}\,\frac{eE_0r_0^3}{L_y^2c^2}
 -
\frac{e^2E_0^2r_0^4}{L_y^2c^2}\,\frac{1}{ m }
\bigg)\,
\frac{1}{(1-\xi)^{1- m }}
- \frac{e^2E_0^2r_0^4}{L_y^2c^2}\,
  \frac{1/ m }{(1-\xi)^{1-2 m }}    \notag\\
 &=
1-\xi 
 -
\bigg(
\frac{\mathcal{E}/e}{L_yc/(er_0)}
\frac{E_0r_0}{L_yc/(er_0)}
 -
\Big(
\frac{E_0r_0}{L_yc/(er_0)}
\Big)^2\,\frac{1}{ m }
\bigg)\,
\frac{1}{(1-\xi)^{1- m }}
 -
\Big(
\frac{E_0r_0}{L_yc/(er_0)}
\Big)^2\,
\frac{1/ m }{(1-\xi)^{1-2 m }}\notag\\
 &=
1-\xi 
 -
\bigg(
\frac{\mathcal{E}}{\mathcal{E}_0}\,
\frac{L_0^2}{L_y^2}
 -
\frac{L_0^2}{L_y^2}\,
\frac{\beta_0^2}{ m }
\bigg)\,
\frac{1}{(1-\xi)^{1- m }}
 -
\frac{L_0^2}{L_y^2}\,
\frac{\beta_0^2}{ m }
\frac{1}{(1-\xi)^{1-2 m }}.
\label{eq:WeakFoc.4}
\end{align}
This equation can be reduced to quadratures but, 
the subsequent integral cannot be performed analytically for 
non-integer $m$ values, and only with difficulty, if at all,
for $m$ values other than 1.
\end{widetext}

\paragraph{\bf Specialization to $m=1$ Planar Orbits.}
The formulas just obtained simplify 
spectacularly for $m$=1. For example, the electric potential 
(defined to vanish for $\xi=0$) is given by
\begin{equation}
V(\xi) = E_0r_0\xi.
\label{eq:meq1.0}
\end{equation}
To obtain the orbit equation we can start from 
Eq.~(\ref{eq:WeakFoc.4}), specialized to the spherical electrode
geometry shown in Fig.~\ref{fig:SphericalElectrodes}.

Eq.~(\ref{eq:WeakFoc.4}) becomes especially simple
for motion in the $y=0$ plane with $m$=1;
\begin{equation}
\frac{d^2\xi}{d\theta^2}
 =
 1 
- \frac{L_0^2}{L_y^2}\,
\frac{\mathcal{E}}{\mathcal{E}_0}
 -
\bigg(
1 - \frac{L_0^2}{L_y^2}\,\beta_0^2
\bigg)\,\xi
 \equiv
-Q_x^2(\xi-\xi_{\rm co}).
\label{eq:WeakFoc.5}
\end{equation}
The last form implicitly defines abbreviations $Q_x$ and $\xi_{\rm co}$.
It also introduces a representation of transverse displacements as the 
superposition of a ``betatron'' part and an ``off-energy'' part.

It can hardly be over-emphasized how simple this
equation of motion is. When expressed in terms of $\xi$ the 
motion is not only simple harmonic, it is simple harmonic
for arbitrarily large oscillation amplitudes. The only 
reservation is that $\xi$ cannot exceed 1, at which point
$r$=$\infty$.

For our application the coefficient $Q_x^2$ is positive,
which means there is horizontal focusing with horizontal
tune $Q_x$ given by 
\begin{equation}
\boxed{
Q_x
 =
\sqrt{1 - \frac{L_0^2}{L_y^2}\,\beta_0^2}.}
\label{eq:WeakFoc.6}
\end{equation}
The off-energy central orbit is given by
\begin{equation}
\boxed{
\xi_{\rm co}
 = 
\frac{1}{Q_x^2}\,
\bigg(
 1 
- \frac{L_0^2}{L_y^2}\,
\frac{\mathcal{E}}{\mathcal{E}_0}
\bigg),
\quad
\xi'_{\rm co}=0.}
\label{eq:WeakFoc.7}
\end{equation}
With $L_y$ and $\mathcal{E}$ allowed to vary,
(especially in RF cavities, but \emph{not within electric bend elements})
the parameters $Q_x$ 
and $x_{\rm co}$ have to be regarded as 
locally, rather than globally, defined.
It is important to notice though, that the parameters entering
the definitions of both $Q_x$ and $\xi_{\rm co}$ are invariants
of any given particle's motion. The quantities ${E}_0$ and
$\beta_0$ are obviously invariant because they are properies
of the on-energy central orbit.  The quantities ${\mathcal{E}}$
and $L_y^2$ are invariant by conservation of energy and angular 
momentum, but they are, in general, 
(slightly) different for different particles.

\paragraph{\bf Particle-specific transfer matrices.}
In practice there will be drift regions, quadrupoles,
RF cavities and
other apparatus in the ring, not to mention artificial
thin quadrupoles inserted within the bend to compensate
for actual field index differing from $m$=1.
Eq.~(\ref{eq:WeakFoc.5})
will be applied only within individual sector bend
elements. But the simplest possible application of
the formulas has a single element bending 
through $2\pi$---which is to say, making up the whole
ring. The $x$ motion described by linearized
treatment is only sinusoidal
for small amplitudes. But, since the denominator is 
a periodic function of $\theta$, $x(\theta)$ has the
same period. 
For the frozen spin value $\beta_x=0.59838$, with
orbits close enough to circular that $L\approx L_0$, 
the tune is $Q_x=0.8012$. (Notice that the parameter $Q$ 
and the Mu\~noz parameter $\kappa$ are identical parameters.) 

Our equations, being fully relativistic, are valid
in the nonrelativistic limit $\beta_0\rightarrow0$.
In the nonrelativistic limit the orbits are simply Keplerian
planetary orbits, which are known to form closed ellipses,
corresponding to $Q_x$=1. So one can say that the 
shift of $Q_x$ away from 1 is a relativistic effect.\footnote{As a
curiosity, one can evaluate the corresponding tune shift
of the orbit of planet Mercury around the sun. We pretend the orbit
is circular (when in fact its eccentricity is $\epsilon=0.21$, 
yielding, for fixed total energy, 
$(L/L_0)^2\approx1/(1-\epsilon^2)\approx1.04$.)  
Mercury's mean orbital velocity is 47.87\,km/s
so $\beta_0=1.596\times10^{-4}$ and 
$Q_x\approx1-\beta_0^2/2=1.27\times10^{-8}$; this represents a
\emph{negative} fractional \emph{advance} of the perihelion
each turn. (Our electrostatic formulation amounts to accounting
for the relativistic increase in inertial mass with no
corresponding increase in gravitational attraction.)
Since its orbital period of 0.241 years, 
Mercury completes 100/0.241=415 revolutions
in 100 years. Meanwhile, its Einsteinian general relativistic 
perihelion advance is 43 arc-sec. Expressed as a fractional
deviation per revolution, this is 
$43/(360\times3600\times415)=7.99\times10^{-8}$. It
appears, therefore, that the perihelion advance is reduced
by about 16\% by the special relativistic inertial increase.
This is roughly consistent with Chandler\cite{EngelkeChandler}.
}

An orbit having initial vertical displacement or slope 
leaves the horizontal (design) plane. It
nevertheless lies in a single plane, and its evolution is
the same as if it were in the design plane. From this
point of view $Q_y=0$. On the other hand, if one insists on 
interpreting non-zero $y$ values as vertical 
betatron oscillations, the vertical tune is $Q_y=1$.
This is an unusual example of ``tune aliasing''. 

For pure horizontal betatron oscillations
\begin{equation}
Q \equiv Q_x = \sqrt{1 - \beta_0^2}.
\label{eq:LattParams.1pp}
\end{equation}
For propagating the radial displacement
$\Delta\xi=\xi-\xi_{\rm co}$,
one can introduce cosine-like
trajectory $C_{\xi}(\theta)$ satisfying $C_{\xi}(0)=1$, $C_{\xi}'(0)=0$ 
and  sine-like trajectory $S_{\xi}(\theta)$ satisfying 
$S_{\xi}(0)=0$, $S_{\xi}'(0)=1$. They are given by
\begin{align}
C_{\xi}(\theta)
 &=
\cos( Q\theta)
\notag\\
C_{\xi}'(\theta)
 &=
- Q\,\sin( Q\theta)                        
\notag\\
S_{\xi}(\theta)
 &=
\frac{\sin( Q\theta)}{ Q}
\notag\\
S_{\xi}'(\theta)
 &=
\cos( Q\theta)
.
\label{eq:Full-Sp.1}
\end{align}
For describing evolution of $(\xi,\xi')$ from its initial
values $(\xi_{\rm in},\xi_{\rm in}')$ at $\theta=0$ to its values at $\theta$
one can use the ``transfer matrix'' defined by
\begin{equation}
{\bf M}_{\xi}\,(\theta)
 =
\begin{pmatrix}
C_{\xi}(\theta)  & S_{\xi}(\theta) \\
C_{\xi}'(\theta) & S_{\xi}'(\theta) 
\end{pmatrix}
 =
\begin{pmatrix}
 C  & S/Q \\
-SQ &  C 
\end{pmatrix},
\label{eq:Full-Sp.2}
\end{equation}
where $C\equiv\cos(Q\theta)$ and $S\equiv\sin(Q\theta)$,
to give
\begin{equation}
\begin{pmatrix} \xi(\theta) \\  \xi'(\theta)\end{pmatrix}
 =
\begin{pmatrix} \xi_{\rm co} \\  0       \end{pmatrix}
 +
{\bf M}_{\xi}(\theta)
\begin{pmatrix} \xi_{\rm in}-\xi_{\rm co} \\ \xi'_{\rm in} \end{pmatrix}.
\label{eq:Full-Sp.3}
\end{equation}
or, spelled out in inhomogeneous form,
\begin{align}
\xi(\theta)
 &=
\xi_{\rm co}
 + 
C\,(\xi_{\rm in}-\xi_{\rm co})
 + 
(S/Q)\,\xi'_{\rm in}, \notag\\
\xi'(\theta)
 &=
\qquad 
-SQ(\xi_{\rm in}-\xi_{\rm co})
 + 
C\xi'_{\rm in}.
\label{eq:Full-Sp.3p}
\end{align}
The subscript $\xi$ on ${\bf M}_{\xi}$ and its components is to serve 
as a reminder that this matrix is particle-specific, and is
therefore not a transfer matrix in the conventional,
linearized, particle-independent, sense. In spite of 
appearing superficially to be linearized, ${\bf M}_{\xi}$ describes 
nonlinear motion, and is exact for all amplitudes.
Eq.~(\ref{eq:Full-Sp.3p}) can be written more compactly as 
\begin{align}
\begin{pmatrix} \xi(\theta) - \xi_{\rm co} \\ \xi'(\theta) \end{pmatrix} 
 &=
{\bf M}_{\xi}
\begin{pmatrix} \xi_{\rm in} - \xi_{\rm co} \\ \xi'_{\rm in} \end{pmatrix} \notag\\
 &=
\begin{pmatrix}
 C  & S/Q \\
-SQ &  C 
\end{pmatrix}
\begin{pmatrix} \xi_{\rm in} - \xi_{\rm co} \\ \xi'_{\rm in} \end{pmatrix}.
\label{eq:Full-Sp.3q}
\end{align}

\begin{widetext}
\paragraph{\bf Propagation Through a Sector Bend}
Starting from known coordinates $(x_{\rm in},x_{\rm in-}')$ 
just preceding the entrance
to a sector bend, where the electric potential vanishes, one
presumably knows $\mathcal{E}$ and, therefore,
\begin{equation}
p^2_{\rm in-} = \mathcal{E}^2/c^2 - m_p^2c^2,\quad
p^2_{z,\rm in-} = \frac{p^2_{\rm in-}}{1+{x'}^2_{\rm in-}},\quad
p_{x,\rm in}
 = 
{x'}_{\rm in-}\,\sqrt{p^2_{z,\rm in-}}.
\label{eq:SectorBend.1}
\end{equation}
(In our hard edge approximation)
just past the entrance, $x$, $\mathcal{E}$, and $p_x$ are unchanged,
but the other dynamic variables have changed. In particular, using
conservation of $p_x$,
\begin{equation}
p^2_{\rm in+}
 = 
\Big(\frac{\mathcal{E}}{c} - \frac{E_0r_0\,x_{\rm in}/c}{r_0+x_{\rm in}}\Big)^2 - m^2_pc^2,\quad
p^2_{z,\rm in+} = p^2_{\rm in+} - p^2_{x,\rm in},\quad
x'_{\rm in+} = \frac{p_{x,\rm in}}{\sqrt{p^2_{z,\rm in+}}}.
\label{eq:SectorBend.2}
\end{equation}
This has determined ${x'}_{\rm in+}$. 
The last expression has been arranged to give not just the magnitude, 
but also the sign of $x'_{\rm in+}$. 
(The argument of the square root can never change sign.) 
Notice that this calculation has included the
(very small, because of near-normal incidence) ``refractive'' bending 
at the interface. Using Eq.~(\ref{eq:WeakFoc.2p}), initial coordinates
$(\xi_{\rm in},\xi_{\rm in}')$, just inside the sector bend
can be obtained;
\begin{equation}
\xi_{\rm in} = \frac{x_{\rm in-}}{r_0+x_{\rm in-}},
\quad
\xi_{\rm in}' = \frac{r^2_0x'_{\rm in+}}{(r_0+x_{\rm in-})^2}.
\label{eq:Full-Sp.4}
\end{equation}
Note also that $L_y$, which is 
proportional to $(r_0+x)p_z$ changes according to
\begin{equation}
\frac{L^2_{y,\rm in+}}{L^2_{y,\rm in-}}
 =
\frac{p_{z,\rm in+}^2}{p_{z,\rm in-}^2},
\label{eq:SectorBend.3}
\end{equation}
with both numerator and denominator known from earlier formulas.
From Eqs.~(\ref{eq:WeakFoc.6}) and (\ref{eq:WeakFoc.7}) it
can be seen that $L_y$ is the only variable parameter  
in the definitions of $Q$ and $x_{\rm co}$; (not counting
changes in $\mathcal{E}$ occurring in RF cavities).
$Q$ and $x_{\rm co}$ can be updated accordingly. With the
updated version of $x_{\rm co}$ one can determine 
$\xi_{\rm in}-x_{\rm co}$ which is needed as an input for
the bend element transfer matrix.

From the known bend angle $\Delta\theta$, cosine and sine factors
$C$ and $S$ can be calculated, and all elements of 
transfer matrix ${\bf M}_{\xi}$ determined. 
Then, using Eq.~(\ref{eq:Full-Sp.3q}), one obtains the components
$(\xi_{\rm out},\xi'_{\rm out})$ just before the exit from the 
sector bend.

To get out of the bend element the steps taken at the input
have to be reversed. 
\end{widetext}

\section{Time of flight and longitudinal phase space dynamics}
The treatment of orbits has been based on positions,
slopes and momenta, with no need for velocities.  To
calculate time of flight, velocity is required. These calculations
will proceed in parallel with the orbit calculations
but they are described separately here to emphasize the
essentially different evolution of position ${\bf r}(\theta)$
and $ct(\theta)$, which is the arrival time (expressed as a length) 
relative to that of the central particle. With all quadrupoles and 
multipoles treated as 
thin in ETEAPOT, the only contributions to time of flight come 
from drifts and electric bend elements. The only essential complication
comes from the dependence of particle speed on position.

Even apart from their different velocities, with all orbits very nearly
parallel, the path length is very nearly the same for all particles.
Yet synchrotron oscillation dynamics depend sensitively of their time
of flight differences. Furthermore, the main reason the spin coherence 
time SCT can be long is the strong tendency for excess spin precession to
cancel over complete synchrotron oscillation cycles. To the extent this
cancellation depends nonlinearly on synchrotron amplitude the SCT is
impaired. These considerations make time of flight calculation
difficult, yet important.

As shown previously, in the $m=1$ Mu\~noz/Pavic treatment the time 
of flight integrals can be evaluated exactly and in closed form.
But the resulting formulas are quite complicated, and numerically
treacherous, because they contain (multiple-valued) inverse tangents
and cotangents. Formulas are given in the ETEAPOT user 
manual\cite{ETEAPOT-expanded}. 

In the paraxial approach (in spite of previously having been shown
to be ``integrable'' for $m=1$) the time of flight integral cannot 
be expressed in closed form. But, after power series expansion, the 
integrals are elementary. The resulting power series is so rapidly 
convergent that accuracy to machine precision is quick. We have
found this approach more robust than the  Mu\~noz Pavic approach
for analysing longitudinal motion.
This formulation is described next.

\paragraph{\bf Kinematic variables within electric bend elements.}
For the $ m =1$, inverse square law case being emphasized, 
the electric field
and electric potential are given by Eqs.~(\ref{eq:WeakFoc.1m}) 
and (\ref{eq:WeakFoc.2m});
\begin{align}
{\bf E}(\xi)
 &= 
-E_0\,(1-\xi)^2\,{\bf\hat r}, \notag\\
V(\xi)
 &=
E_0r_0\xi.
\label{eq:Kinematic.1}
\end{align}
The latter equation, along with Eq.~(\ref{eq:Full-Sp.3p}),
permits the potential energy to be expressed in terms
of $\theta$ and initial conditions.
\begin{equation}
V(\theta)
 =
E_0r_0\,
\Big(
\xi_{\rm co}
 + 
(\xi_{\rm in}-\xi_{\rm co})\,\cos Q\theta
 + 
\frac{\xi'_{\rm in}}{Q}\,\sin Q\theta.
\Big)
\label{eq:Kinematic.3pm}
\end{equation}
Then $\gamma(\theta)$ is obtained from
\begin{equation}
\gamma(\theta)
 =
\frac{\mathcal{E}}{m_pc^2}
 -
\frac{E_0r_0}{m_pc^2/e}\,
\Big(
\xi_{\rm co}
 + 
(\xi_{\rm in}-\xi_{\rm co})\,\cos Q\theta
 + 
\frac{\xi'_{\rm in}}{Q}\,\sin Q\theta
\Big).
\label{eq:Kinematic.3pmp}
\end{equation}
From this formula one can obtain $\beta(\theta)$ using
\begin{equation}
\beta(\theta)
 =
\sqrt{1 - 1/\gamma^2}.
\label{eq:Kinematic.3p}
\end{equation}
From the $y$-component of Eq.~(\ref{eq:Cylinder.4}) we also have,
for motion in the horizontal plane,
\begin{equation}
\frac{d\theta}{dt}
 =
\frac{-L_y}{m_pr^2\gamma}.
\label{eq:Kinematic.4}
\end{equation}

\noindent
{\bf Warning:} 
Note that, with right-handed Frenet coordinates, and clockwise orbit (which
we now assume) $L_y$ is negative. This accounts for the negative sign
in Eq.~(\ref{eq:Kinematic.4}). 
The right hand side of this equation can now be expressed in 
terms of $\theta$, invariants and initial conditions.
The negative sign ($-L_y$) is specific to clockwise orbits, with
$\theta$ increasing along the orbit. There seems to be no universally
accepted sign convention distinguishing clockwise and 
counter-clockwise beam evolution in storage rings.
This sign ambiguity causes the tracked time of flight variable $ct$ 
to also be ambiguous. The appropriate RF phase is then, similarly, 
ambiguous.  

\paragraph{\bf $\xi$ evolution and convergence estimates.}
Using Eq.~(\ref{eq:WeakFoc.2p}), initial conditions $(\xi_0,\xi_0')$ 
have been expressed in terms of initial $x$ conditions;
\begin{equation}
\xi = \frac{x}{r_0+x},
\quad
\xi' = \frac{r^2_0x'}{(r_0+x)^2}.
\label{eq:xi2x.0}
\end{equation}
The evolution of $x$ is obtained using Eq.~(\ref{eq:Full-Sp.3p}), 
whose upper equation yields
\begin{equation}
\xi
 =
\xi_{\rm co}
 + 
(\xi_{\rm in}-\xi_{\rm co})\,\cos Q\theta
 + 
\frac{\xi'_{\rm in}}{Q}\,\sin Q\theta.
\label{eq:xi2x.1}
\end{equation}
This formula can be used to give $\xi$ as a function
of $\theta$, for example when calculating times of flight. 
Then 
\begin{equation}
x(\theta) = \frac{r_0\xi(\theta)}{1-\xi(\theta)},
\quad\hbox{and}\quad
r(\theta) = r_0 + x(\theta).
\label{eq:xi2x.1p}
\end{equation}
For the proton EDM experiment the value of radius $r_0$ will be, 
say, 40\,m. An initial value $x_0=1\,$m would be unrealistically
large. It is better therefore to use centimeter units. Then
$r_0=4000\,$cm, and the
initial displacement of a cosine-like trajectory is 1\,cm. With
the electrode spacing being 3\,cm, a unit-amplitude,
cosine-like trajectory happens to have something like a maximal 
amplitude. Dropping terms of one higher order, for example from
quadrupole order to sextupole order, therefore amounts to making an 
error of about one part in 4000. An error of this magnitude is likely
to be unimportant unless some resonance causes its repeated
effect to accumulate constructively over many turns.

\paragraph{\bf Time of flight through bends.}
A formula for the time of flight has already been given in 
Eq.~(\ref{eq:RevPeriod.1}). Since this form is integrable,
it gives the exact flight time in closed analytic form. However 
it has the disadvantage that the angle $\theta_0$ 
(which is the angle to ``perihelion'') 
is hard to determine unambiguously, because it is multiple-valued,
and because perihelion is indeterminate for circular orbits.
The most important circular orbit is the design central orbit
and, depending on elements encountered in the ring, the local
angle to perihelion can vary erratically.   

It is useful, therefore, to have an alternate, highly accurate, 
though not necessarily analytically exact, form, to accompany 
the Mu\'noz form
developed earlier. From Eq.~(\ref{eq:Kinematic.4}) the time of 
flight $dt$, expressed as a distance $d(ct)$ is given by
\begin{equation}
\frac{dct}{d\theta}
 =
\frac{m_pc^2r^2}{-L_yc}\,\gamma
 =
\frac{1}{-L_yc/e}\,\frac{r_0^2}{(1-\xi)^2}\,
(
\mathcal{E}/e
 -
E_0r_0\xi
)
.
\label{eq:TimeOfFlight.1}
\end{equation}
where
\begin{equation}
\boxed{
\xi
 =
\xi_{\rm co}
 + 
(\xi_{\rm in}-\xi_{\rm co})\,\cos Q\theta
 + 
\frac{\xi'_{\rm in}}{Q}\,\sin Q\theta.}
\label{eq:TimeOfFlight.1p}
\end{equation}
Note that the variable factor $r^2$ has been replaced
using Eq.~(\ref{eq:WeakFoc.2pp}) and that the parameters $Q$ and 
$\xi_{\rm co}$, though invariant, are particle-dependent as in 
Eqs.~(\ref{eq:WeakFoc.6}) 
and (\ref{eq:WeakFoc.7}). (The proliferation of ``/e'' 
factors is for convenience in converting electron-volts
to MKS energy units.) 
Formula~(\ref{eq:TimeOfFlight.1}) can be checked on the design 
orbit where $\xi=0$, using $L_0c=-m_pc^2\gamma_0\beta_0r_0$;
\begin{equation}
\frac{dct_0}{d\theta}
 =
\frac{1}{\gamma_0\beta_0r_0m_pc^2/e}\,r^2_0\,(\gamma_0m_pc^2/e)
 =
\frac{r_0}{\beta_0},
\label{eq:TimeOfFlight.3}
\end{equation}
which is correct. 
The time of flight through an element with angle $\Delta\theta$ is
given by
\begin{equation}
ct 
 =
\int_0^{\Delta\theta}\,
\frac{r_0^2(\mathcal{E}/e - E_0r_0\xi)}{-L_yc/e}\,\frac{1}{(1-\xi)^2}\,
d\theta,
\label{eq:TimeOfFlight.3p}
\end{equation}
(where, for clockwise motion, $-L_y$ is positive).
Over the corresponding sector the time of flight of the design particle is 
\begin{equation}
ct_0
 =
\int_0^{\Delta\theta}\,
\frac{r_0}{\beta_0}
d\theta.
\label{eq:TimeOfFlight.3q}
\end{equation}
The increment in time relative to the design particle is
\begin{equation}
ct - ct_0
 =
\int_0^{\Delta\theta}\,
\bigg(
\frac{r_0^2(\mathcal{E}/e - E_0r_0\xi)}{-L_yc/e}\,\frac{1}{(1-\xi)^2}\,
 -
\frac{r_0}{\beta_0}
\bigg)\,d\theta,
\label{eq:TimeOfFlight.3r}
\end{equation}
The integrand can be expanded in a series in $\xi$;
\begin{equation}
\frac{r_0^2(\mathcal{E}/e - E_0r_0\xi)}{-L_yc/e}\,\frac{1}{(1-\xi)^2}\,
 -
\frac{r_0}{\beta_0}
 =
A_0 + A_1\xi + A_2\xi^2 + A_3\xi^3 + \dots,
\label{eq:TimeOfFlight.3s}
\end{equation}
where
\begin{align}
A_0 &= -\frac{r_0^2\mathcal{E}/e}{L_yc/e} - \frac{r_0}{\beta_0},\quad 
A_1  = -\frac{2 r_0^2 \mathcal{E}/e}{L_yc/e} + \frac{r_0^3 E_0}{L_yc/e}, \notag \\ 
A_2 &= -\frac{3 r_0^2 \mathcal{E}/e}{L_yc/e} + \frac{2 r_0^3 E_0}{L_yc/e},\quad 
A_3  = -\frac{4 r_0^2 \mathcal{E}/e}{L_yc/e} + \frac{3 r_0^3 E_0}{L_yc/e},\notag \\   
A_4 &= -\frac{5 r_0^2 \mathcal{E}/e}{L_yc/e} + \frac{4 r_0^3 E_0}{L_yc/e},\quad  
A_5  = -\frac{6 r_0^2 \mathcal{E}/e}{L_yc/e} + \frac{5 r_0^3 E_0}{L_yc/e},
\dots.  
\label{eq:TimeOfFlight.5}
\end{align}
All the required integrals (over the range from 0 to $\Delta\theta$, which
is the bend angle) are elementary---the integrands are integer powers of 
$\sin Q\theta$ or $\cos Q\theta$, and the series in Eq.~(\ref{eq:TimeOfFlight.3s}) 
is rapidly convergent. It is important to remember that quantities, in this case
$L_y$, which are discontinuous in the transition from just outside to just
inside a bend element, have to be updated to the value \emph{inside} before evaluating
these $A_i$ coefficients. 

Because $\xi^2$ is necessarily positive while $\xi$
can have either sign, the ``quadratic'' term proportional to $\xi^2$ may compete 
with the preceding ``linear'' term proportional to $\xi$, which can have either 
sign and will tend to cancel on the average. As has been stated repeatedly, 
since the coefficients are particle-dependent, this formula has to be
performed individually for every particle at every bend, 

\paragraph{\bf Time of flight through straight sections.}
Different particles also have different flight times through drift
sections or multipole elements of length $\ell$.
The path length excess is approximately
\begin{equation}
c(t-t_0)_{\rm path}
 =
\frac{{x'}^2\ell^2+{y'}^2\ell^2}{2\ell\beta_0}
 =
\frac{({x'}^2+{y'}^2)\ell}{2\beta_0},
\label{eq:TimeOfFlight.6}
\end{equation}
and the off-speed correction is
\begin{equation}
c(t-t_0)_{\rm vel.}
 =
\ell\,\bigg(\frac{1}{\beta} - \frac{1}{\beta_0}\bigg).
\label{eq:TimeOfFlight.7}
\end{equation}
The total effect is 
\begin{equation}
c(t-t_0)_{\rm straight}
 =
\frac{\ell}{\beta_0}
\bigg( 
\frac{\beta_0}{\beta} - 1 + \frac{{x'}^2+{y'}^2}{2}
\bigg),
\label{eq:TimeOfFlight.8}
\end{equation}
or, more directly,
\begin{equation}
c(t-t_0)_{\rm straight}
 =
\frac{\ell\sqrt{1+{x'}^2}}{pc/\mathcal{E}} - \frac{\ell}{\beta_0}.
\label{eq:TimeOfFlight.8p}
\end{equation}

\paragraph{\bf Time of flight due to vertical oscillation.}
For bends only the projection of the orbit onto the horizontal plane has been
found so far.  It remains to evaluate the time of flight ascribable to
vertical betatron motion. We assume there is no direct coupling between
$x$ and $y$. We also neglect the effect of the small speed variations
accompanying horizontal oscillations. Also neglecting the small bending
in sufficiently-finely sliced bend elements, we treat them as straight.
With these approximations the bend element is treated as a drift as
far as vertical oscillations are concerned. The off-velocity effect has
already been included in the horizontal calculation. 
Copying from 
Eq.~(\ref{eq:TimeOfFlight.8}),
\begin{equation}
c(t-t_0)_{\rm vert.}
 =
\frac{\ell\,{y'}^2}{2\beta_0}.
\label{eq:TimeOfFlight.9}
\end{equation}

\section{Lumped correction for field index deviation}
Analytic propagation formulas have been given for $1/r^2$
radial dependence of the electric field. 
The actual radial dependence will deviate from this, with $m$, 
the deviant field index, having been defined in 
Eq.~(\ref{eq:WeakFoc.1m}). $E\sim1/r^{1+m}$, so $m=1$ for
the Coulomb's law field dependence. For this, the ``spherical'' 
case, the orbit equations have been solved 
in closed form. 

The ETEAPOT strategy, even for $m\ne1$, is to treat sector bends 
as \emph{thick} elements with orbits given by the analytic 
$m=1$ formulas. The $m\ne1$ case is handled by inserting
zero thickness ``artificial quadrupoles'' of appropriate strength. 
As with TEAPOT, this approximation becomes arbitrarily good in the
thin slicing limit. Though the idealized model differs from
the physical apparatus, the orbit description within the 
idealized model is exact, and 
hence symplectic. For too coarse slicing, even though the orbit 
deviates from the ``true'' orbit, it remains ``exact'' (for a
lattice with that slicing).
One blames the
inaccuracy on the fact that the lattice model deviates from the
true lattice, not on the fact that the equations are being
solved approximately. By varying the slicing one can judge
from the limiting behavior what degree of slicing is appropriate.

In a magnetic fields there is ``geometric'' horizontal focusing
even in a uniform field. (For example a cyclotron orbit
with non-vanishing initial slope returns to its starting point
after one full turn---its horizontal tune therefore being 1.0.)
A deviant field index $m$ introduces further horizontal and vertical
focusing terms of equal magnitude, but of opposite sign, into 
the focusing equations

The corresponding focusing strength coefficients in electric 
fields are\cite{Wollnik}
\begin{align}
K^{(m)}_{x,1}
 &=
\bigg(
\frac{1}{r^2_0} + \frac{1}{\gamma_0^2r^2_0}
\bigg)
 - \frac{m}{r^2_0} ,
\notag \\
K^{(m)}_{y,1}
 &=
\frac{m}{r^2_0}.
\label{eq:lumped.2}
\end{align}
Note that for $m=0$ (the ``cylindrical'' case) there is no
vertical focusing. With the electric field being independent of $y$ 
for cylindrical electrodes, this is obviously correct.
(The subscript ``1'' indicates ``quadrupole'' order; a more 
careful treatment brings in higher order corrections.)
The parentheses segregate the geometric focusing 
(which is independent of $m$) from the gradient focusing 
(which vanishes for $1/r$ field variation).
Re-writing these equations for our spherical $m=1$ case;
\begin{align}
K^{\rm sph.}_{x,1}
 &=
\bigg(
\frac{1}{r^2_0} + \frac{1}{\gamma_0^2r^2_0}
\bigg)
 - \frac{1}{r^2_0} ,
\notag \\
K^{\rm sph.}_{y,1}
 &=
\frac{1}{r^2_0}.
\label{eq:lumped.2p}
\end{align}
With our sign convention, positive $K$-value corresponds to
\emph{focusing}. Hence, for example, there is already ``one unit'' 
(in units of $1/r_0^2$) of vertical
\emph{focusing} in the Coulomb case. In the ETEAPOT formalism the focusing
given by Eqs.~(\ref{eq:lumped.2p}) is already implicitly included
since the particle orbits are being tracked exactly. 

For modeling the focusing for $m$ values other than $m$=1
the focusing coefficients we need are given by
Eqs.~(\ref{eq:lumped.2}). The geometric focusing terms
already match but we have to make up the differences
in gradient focusing:
\begin{align}
\Delta K^{(m)}_{x,1}
 &=
K^{(m)}_{x,1} - K^{\rm sph.}_{x,1}
 =
\frac{-m+1}{r^2_0},
\notag \\
\Delta K^{(m)}_{y,1}
 &=
K^{(m)}_{y,1} - K^{\rm sph.}_{y,1}
  =
\frac{m-1}{r^2_0}.
\label{eq:lumped.2q}
\end{align}
These are the coefficients of the distributed focusing we have
to apply to ``convert'' the Coulomb formulas to model the
actual $m$ value. The pure $m$=1 Coulomb field itself 
provides ``one unit'' of vertical focusing. 
The strengths of the quads we insert artificially are therefore
the sum of two terms; one cancels the Coulomb defocusing
handled by the ETEAPOT formalism; the other superimposes 
focusing proportional to $m$, equal but opposite horizontal
and vertical. 

The artificial quadrupoles have zero length, but their length-strength
product has to be matched to the ``field integral'' corresponding to 
length $L_{\rm bend}$ of the bending slice being compensated.

Before continuing
with the treatment for $m\ne0$ it is important to remember the
discontinuous increments to $\mathcal{E}$ as a particle enters
or leaves a bending element. The discontinuity is equal
(in magnitude) to the change in potential energy. When
correcting the focusing by artificial quadrupoles we have
to decide whether the artificial quadrupoles are ``inside''
or ``outside'', since the actual deflections will be
different in the two cases.
Since the difference is
quadratic in transverse displacement, the difference would 
be called 
a ``sextupole'' effect. By treating the artificial quad
as ``inside'', which we do (because we decline to include 
longitudinal drift at zero potential), we avoid introducing an 
artificial sextupole effect. 

Whatever criteria for slicing quadrupoles that have been used
for magnetic rings, the same criteria apply here. The default
slicing in ETEAPOT treats a thick bend as a half bend, a single
kick at the center, and then another half bend. For example,
with $m=-1.3$, the central quad is vertically defocusing
with inverse focal length $q=1/f=2.3\ell/r_0^2$, where $\ell$ is
the arc length through the full bend. 

Numerically, with $r_0=40\,$m and $\ell$ equal to, say, 
$16\,$m, the ratio of focal length to element length is
$(40/16)^2/2.3\approx2.7$. With element length small compared
to focal length, this suggests that the
compensation will be fairly good even with such coarse slicing.
The entire EDM ring, with its 8 full cells, would then be 
represented by 16 half-bends. In practice one will 
probably choose less coarse slicing.

The transfer matrices for the thin effective quadrupole are 
\begin{align}
{\bf K_x^{(m)}}(\Delta\theta)
 &=
\begin{pmatrix}          1               &  0 \\
               ( m-1)\Delta\theta/r_0  &  1
\end{pmatrix},              \notag\\
{\bf K_y^{(m)}}(\Delta\theta)
 &=
\begin{pmatrix}          1               &  0 \\
                (-m+1)\Delta\theta/r_0 &  1
\end{pmatrix},             
\label{eq:lumped.3}
\end{align}
For $m=1$ these kick matrices 
reduce to identity matrices.

With linearized  evolution through half bend formally represented 
by ``transfer matrix'' ${\bf B_{\Delta\theta/2}^{(1)}}$,
bend/kick/bend evolution through bend angle $\Delta\theta$ in 
an element with field index $m$ is then described by the matrix
\begin{equation}
{\bf M_{\Delta\theta}^{(m)}}
 =
{\bf B_{\Delta\theta/2}^{(1)}}
{\bf K^{(m)}}(\Delta\theta)
{\bf B_{\Delta\theta/2}^{(1)}}.
\label{eq:BendKickBend.1}
\end{equation}

\section{Spin tracking}
\paragraph{\bf Spin tracking through thick elements.}
The second main new feature implemented in ETEAPOT is spin evolution.
To leading approximation spin precession occurs in central force,
inverse square law, force field regions. With this assumption
the orbit through a bend element of any individual particle lies in
a single fixed plane. 

Each individual particle's spin vector can be decomposed
into a (conserved) component $\tilde s_{\perp}$, normal to the bend plane, 
and a (precessing) 2-component vector ${\bf \tilde s_{\parallel}}$, 
lying in the bend plane.

To lowest approximation hard edge bends are assumed, though a refractive
deflection accompanying the change in electric potential is included.
To a next approximation the fringe field electric potential is represented 
as a linear ramp. In this region, because the particle speed is necessarily 
non-magic, there is a noticeable difference in precession rates of
spin and momentum. Furthermore the precession error at entrance and
exit add constructively. It is assumed that the paths through the entrance 
and exit fringe fields lie in the same plane as in the bend interior.

As always in ETEAPOT, field deviations from radial inverse square law are 
modeled by artificial quadrupole ``kicks''.

Quadrupoles, whether real or artificial (as well as all other multipoles) 
are treated as thin. This includes the approximation that the electric 
potential is independent of position, both transversely and longitudinally
in the element. Because the element thickness is treated as zero, there 
is no further time of flight error caused by the neglect of speed 
changes as the particle passes through the multipole.

Superimposed multipoles are therefore \emph{exactly} 
superimposed which means that spin precession are modeled by successive 
rotations, with no dependence on the order in which they are applied.
All such spin rotations are concatenated explicitly into a single 
(near-identity) precession matrix. For improved numerical precision
all such concatenations are performed analytically and the result expressed
as the exact sum of an identity matrix and a deviation matrix. This 
circumvents the problem of large, approximately canceling precessions 
and avoids spurious non-commutative geometric precessions.

Quadrupoles too thick to be validly treated as thin, are sliced, with
regions between the sliced thin quadrupoles treated as drifts.

\begin{figure}[h]
\centering
\includegraphics[scale=0.4]{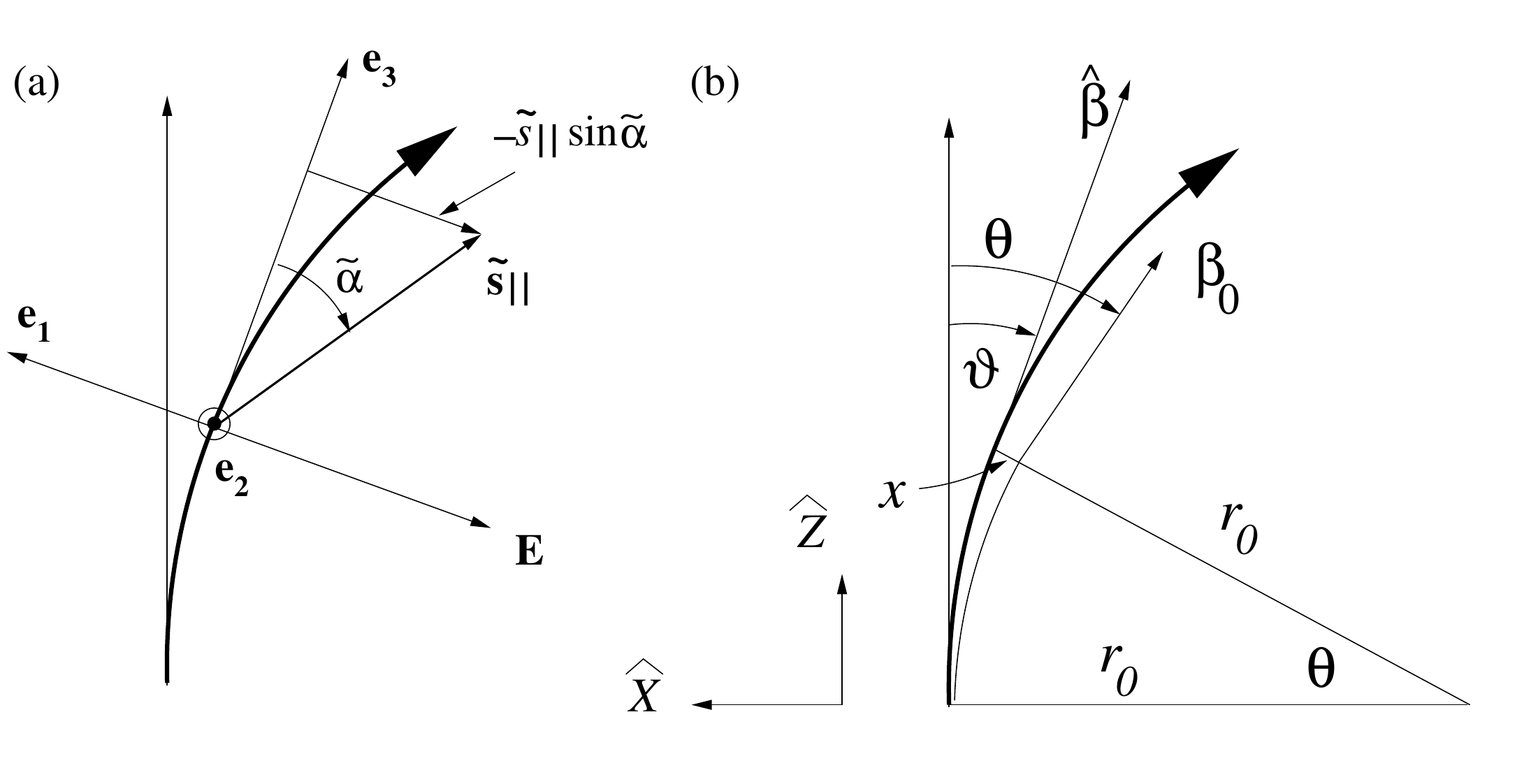}
\caption{\label{fig:Precess}
(a) In the bend plane the
spin vector ${\bf s}$ has precessed through angle
$\tilde \alpha$ away from its nominal direction along the
proton's velocity. (Remember that different particles
have different bend planes.)
(b) Projection of figure~(a) onto the laboratory horizontal
plane. The projected longitudinal axis is shown
coinciding with the laboratory longitudinal axis, even
if this is not exactly valid. $x$ is the deviation of 
the (bold face) particle orbit from the (pale face) design orbit. 
If the bend plane coincides with the design bend plane (as is 
always approximately the case) $\hat{\pmb \beta}_0$ and 
${\bf\hat z}$ are identical. $\theta$ is the reference
particle deviation angle from longitudinal and $\vartheta$
is the tracked particle deviation angle from longitudinal. 
On the average $\theta$ and $\vartheta$ are the same, but
betatron oscillations cause them to differ on a turn by turn
basis, and also to make the instantaneous bend plane not
quite horizontal.}
\end{figure}

\paragraph{\bf Spin representation.}
Laboratory frame spin components are  $(s_x,s_y,s_z)$.
Bend frame spin components $(\tilde s_x,\tilde s_y,\tilde s_z)$, are 
illustrated on the left in Figure~\ref{fig:Precess}, which projects the 
spin vector onto the bend plane.
\begin{equation}
\vec{ \tilde S} =
\begin{pmatrix} \tilde s_x \\ \tilde s_y \\ \tilde s_z \end{pmatrix}
 =
\begin{pmatrix}
 -\tilde s_{\parallel}\sin\tilde\alpha \\ 
   \tilde s_{\perp} \\ 
   \tilde s_{\parallel}\cos\tilde\alpha \end{pmatrix}.
\label{eq:SpinPrecess.0m}
\end{equation}
As with the orbit equation, the BMT equation is exactly solvable only
because the orbit stays in a single bend plane. For particles in realistic 
beams this is very nearly, but not exactly, the local $(x,y)$ central 
design plane. 
\paragraph{\bf Transforming spin components from Lab frame to 
bend frame (and back again).}
From orbit tracking to some current location in the ring 
one knows the laboratory frame
position, ${\bf r}$, and momentum, {\bf p}, 
and hence the angular momentum
${\bf L}={\bf r}\times {\bf p}$.
One also knows the spin vector ${\bf s}$;
\begin{align}
{\bf r} &= r_x{\bf \hat x} + r_y{\bf \hat y} + r_z{\bf \hat z}, \notag\\ 
{\bf p} &= p_x{\bf \hat x} + p_y{\bf \hat y} + p_z{\bf \hat z}, \notag\\ 
{\bf L} &= L_x{\bf \hat x} + L_y{\bf \hat y} + L_z{\bf \hat z}, \notag\\ 
{\bf s} &= s_x{\bf \hat x} + s_y{\bf \hat y} + s_z{\bf \hat z}. 
\label{eq:SpinRot.1}
\end{align}
All of these quantities refer to a point infinitesimally past the bend 
entrance, after refractive correction, but not yet having included any 
entrance fringe field effect. Any spin precession in the interior 
(including fringe fields) of a bend element occurs in a single
bend plane. To exploit this reduction from 3D to 2D
it is first necessary to obtain the
spin components in an orthonormal frame having its ``y'' axis perpendicular 
to the plane and its ``z'' coordinate tangential to the orbit. 
The purpose of this section is to document this transformation.
In the (always excellent) paraxial approximation, and for near-magic 
particle velocity, this transformation will be close to identity.

We can establish an orthonormal, right-handed basis triad with axis~3
parallel to ${\bf p}$ and axis~2 parallel to ${\bf -L}$ (where the
negative sign is appropriate for clockwise orbits);
\begin{align}
{\bf e_3}
 &= 
  \frac{p_x}{p}\,{\bf \hat x} 
+ \frac{p_y}{p}\,{\bf \hat y} 
+ \frac{p_z}{p}\,{\bf \hat z}, \notag\\
{\bf e_2} &= \frac{{\bf r}\times {\bf p}}{-L}, \notag\\
{\bf e_1} &= {\bf e_2}\times{\bf e_3}.
\label{eq:SpinRot.2}
\end{align}
These equations can be re-expressed formally, with all coefficients known, as
\begin{align}
{\bf e_1}
 &= 
a_{11}{\bf\hat x}+a_{12}{\bf\hat y}+a_{13}{\bf\hat z}\notag\\
{\bf e_2}
 &= 
a_{21}{\bf\hat x}+a_{22}{\bf\hat y}+a_{23}{\bf\hat z}\notag\\
{\bf e_3}
 &= 
a_{31}{\bf\hat x}+a_{32}{\bf\hat y}+a_{33}{\bf\hat z}.
\label{eq:SpinRot.3}
\end{align}
The vector ${\bf s}$ can be expanded as
\begin{align}
{\bf s}
 &=
\tilde s_1{\bf e_1} + \tilde s_2{\bf e_2} + \tilde s_3{\bf e_3} \notag\\
 &= \tilde s_1( a_{11}{\bf\hat x}+a_{12}{\bf\hat y}+a_{13}{\bf\hat z}) + \dots \notag\\
 &= (a_{11}\tilde s_1 + a_{21}\tilde s_2 + a_{31}\tilde s_3)\,{\bf\hat x} + \dots.
\label{eq:SpinRot.4}
\end{align}
The final relation can be expressed in matrix form as
\begin{equation}
\begin{pmatrix} s_x \\ s_y \\ s_z \end{pmatrix}
 =
{\bf R}
\begin{pmatrix} \tilde s_1 \\  \tilde s_2 \\  \tilde s_3 \end{pmatrix},
\label{eq:SpinRot.5}
\end{equation}
where ${\bf R}$ is an orthogonal matrix,
\begin{equation}
{\bf R}
 =
\begin{pmatrix} a_{11} & a_{21} &  a_{31}  \\  
                a_{12} & a_{22} &  a_{32}  \\  
                a_{13} & a_{23} &  a_{33}  
\end{pmatrix}.
\label{eq:SpinRot.6}
\end{equation}
(Aside: the magnitude $|{\rm det}{\bf R}|$ of the determinant of 
${\bf R}$ is necessarily 1, but the actual value is $\pm1$.
This sign correlates with the clockwise/counterclockwise
orbit ambiguity.)

Because ${\bf R}$ is orthogonal, ${\bf R}^{-1}={\bf R}^T$ and  
Eq.~(\ref{eq:SpinRot.5}) can be inverted to give 
\begin{equation}
\begin{pmatrix} \tilde s_1 \\  \tilde s_2 \\  \tilde s_3 \end{pmatrix}
 =
\begin{pmatrix} a_{11} & a_{12} &  a_{13}  \\  
                a_{21} & a_{22} &  a_{23}  \\  
                a_{31} & a_{32} &  a_{33}
\end{pmatrix} \,
\begin{pmatrix} s_x \\ s_y \\ s_z \end{pmatrix}.
\label{eq:SpinRot.7}
\end{equation}
This yields the spin components in the bend frame. Their propagation
through the bend is described below.

After the bend plane spin components have been 
updated at the output of the bend element it is
necessary to transform them back to the (local) 
laboratory frame. This entails repeating the preceeding formulas 
starting with Eqs~(\ref{eq:SpinRot.1}), but with ${\bf r}$, ${\bf p}$
and ${\bf L}$ having been evaluated (in local laboratory coordinates)
just inside the output face of the bend element. In other words
all of the quantities in Eqs~(\ref{eq:SpinRot.1}) will now refer
to a point just before the bend exit. Then, following 
Eqs.(\ref{eq:SpinRot.2}) through (\ref{eq:SpinRot.6})
produces rotation matrix ${\bf R}$ which is now applicable at
the bend output. Finally, modifying Eq.~(\ref{eq:SpinRot.7}) 
appropriately, the transformation back to laboratory components 

\begin{equation}
\begin{pmatrix} s_x \\ s_y \\ s_z \end{pmatrix}
 =
\begin{pmatrix} a_{11} & a_{12} &  a_{13}  \\  
                a_{21} & a_{22} &  a_{23}  \\  
                a_{31} & a_{32} &  a_{33}
\end{pmatrix}^T \,
\begin{pmatrix} \tilde s_1 \\  \tilde s_2 \\  \tilde s_3 \end{pmatrix}
.
\label{eq:SpinRot.7p}
\end{equation}

\paragraph{\bf Exact solution of the BMT equation.}
We introduce, at least temporarily, the term 
`''inverse square law bend'' to characterize a bend
having field index $m=1$, which is the case being treated in our
2D formalism. In this case the orbit stays in a single plane. Also,
in this frame any precession of the spin is purely around an 
axis normal to the plane. Obtaining the initial values of the
spin components in this frame was described in the previous section. 

In the bend plane the orbit lies in a single plane.
Superficially this may suggest we are accounting
only for horizontal betatron oscillations and neglecting vertical 
betatron oscillations. In fact, however, the
ETEAPOT treatment accounts for arbitrary betatron and synchrotron
motion by assigning different ``wobbling planes'' to each
individual particle. Even allowing for vertical betatron motion
these frames are all very nearly parallel to the global 
horizontal design frame of the ring.
For the 2D evolution through electric bend elements 
in ETEAPOT, any betatron oscillations actually present for
a particular particle are folded into the determination of
its particle-specific orbit plane, and the initial coordinates in 
this plane.

As shown in Figure~\ref{fig:Precess}, the initial spin vector is
\begin{equation}
{\bf \tilde s}
 = 
 -\tilde s_{\parallel}\sin\tilde\alpha\,{\bf\hat x}
 +\tilde s_y{\bf\hat y} 
 +\tilde s_{\parallel}\cos\tilde\alpha\,{\bf\hat z}.
\label{eq:SpinPrecess.9}
\end{equation}
Here $\tilde s_y{\bf\hat y}$ is the out-of-plane component of ${\bf \tilde s}$,
$\tilde s_{\parallel}$ is the magnitude of the in-plane projection of ${\bf \tilde s}$, 
and $\tilde\alpha$
is the angle between the projection of ${\bf \tilde s}$ onto the plane
and the tangent vector to the orbit.

Jackson's\cite{Jackson} Eq.~(11.171) gives the rate of change 
in an electric field ${\bf E}$, of the longitudinal spin component as
\begin{equation}
\frac{d}{dt}\,
({\bf\hat\beta\cdot s})
 =
-\frac{e}{m_pc}\,
({\bf s_{\perp,J}\cdot E})
\bigg(
\frac{g\beta}{2} - \frac{1}{\beta}
\bigg).
\label{eq:SpinPrecess.1}
\end{equation}
(Note that Jackson's ${\bf s}_{\perp,J}$ is the component perpendicular
to the tangent to the orbit \emph{not} to the orbit plane.)
Substituting from Eq.~(\ref{eq:SpinPrecess.9}) the 
equation becomes
\begin{equation}
\frac{d}{dt}\,
(\tilde s_{\parallel}\cos\tilde\alpha)
 =
-\frac{e}{m_pc}\,
(\tilde s_{\parallel}\sin\tilde\alpha\,E)\,
\bigg(
\frac{g\beta}{2} - \frac{1}{\beta}
\bigg).
\label{eq:SpinPrecess.1mp}
\end{equation}
With the orbit confined to a plane,
any precession occurs about the normal to the plane,
conserving $\tilde s_y$. Since the magnitude of ${\bf \tilde s}$ is
conserved it follows that the magnitude $\tilde s_{\parallel}$
is also conserved. This allows $\tilde s_{\parallel}$ to be 
treated as constant in Eq.~(\ref{eq:SpinPrecess.1mp}).
Then Eq.~(\ref{eq:SpinPrecess.1mp}) reduces to
\begin{equation}
\frac{d\tilde\alpha}{dt}\,
 =
\frac{eE}{m_pc}\,
\bigg(
\frac{g\beta}{2} - \frac{1}{\beta}
\bigg).
\label{eq:SpinPrecess.2}
\end{equation}
This is undoubtedly a fairly good approximate equation in any
more-or-less constant electric field, but it is \emph{exact only for
the $m=1$ Keplerian electric field}, which is the only field in 
which each orbit stays in its own fixed bend plane.  In fact, the
derivation is not quite valid even for our $m=1$ case---though
the design orbit is circular, the betatron orbits are slightly
elliptical. This violates our assumption that orbit and electric 
field are exactly orthogonal. Neglecting this error amounts to dropping 
a term from the RHS of Eq.~(\ref{eq:SpinPrecess.1}). This term is
down by four orders of magnitude compared to the calculated
value. Furthermore this term would average to zero (over many betatron 
cycles) except for a possible non-zero (quadratic in bend angle) 
commutation geometric phase error. Such an error would be expected to be 
down by another four orders of magnitude. Furthermore, the importance 
of this ``error'' can be investigated, and reduced, by finer slicing of 
the bend element.

Meanwhile the velocity vector itself has precessed by angle 
$\vartheta$ relative to a direction fixed in the laboratory. 
Note that this angle $\vartheta$, the angle of the 
individual particle's orbit is approximately, but not exactly 
equal to the angle $\theta$ of the design orbit.

In the ETEAPOT treatment each particle in an electric bend
element evolves in its own bend plane. 
${\bf \tilde s}_{\parallel}$ is the component in this plane of the total
spin vector. At every entrance to an electic bend ${\bf \tilde s}_{\parallel}$
has to be calculated from the known laboratory frame description
of ${\bf s}$, which also has to be updated as the particle exits
the bend. (Ideally, in an EDM storage ring experiment any out-of-plane
component of ${\bf s}$ would be evidence of non-vanishing electric dipole
moment.) 

The precession rate of $\vartheta$ is governed by the equation
\begin{equation}
\frac{d\vartheta}{dt}
 =
\frac{d}{dt}\,\bigg(\frac{s}{r}\bigg)
 =
\frac{eE}{p}.
\label{eq:SpinPrecess.3}
\end{equation}
where the curvature is $1/r=eE/(v p)$ and
(just in this equation) $s$ temporarily stands for arc
length along the orbit. Dividing Eq.~(\ref{eq:SpinPrecess.2})
by Eq.~(\ref{eq:SpinPrecess.3}) and using $pc=m_pc^2\gamma\beta$,
\begin{equation}
\boxed{
\frac{d\tilde\alpha}{d\theta}
 =
\bigg(
\frac{g}{2} - 1
\bigg)\,\gamma
 -\frac{g/2}{\gamma}
.
\label{eq:SpinPrecess.4}
}
\end{equation}
In this step we have also surrepticiously made the
replacement $\vartheta\rightarrow\theta$. Though
these angles are not the same, over arbitrarily long
times they necessarily advance at the same rate. In any 
case the error in equating $\vartheta$ to $\theta$
becomes progressively more valid in the fine-slicing 
limit, as the orbit is more nearly approximated by
straight line segments. Explicitly the bend frame precession 
advance is the sum of two definite integrals
\begin{equation}
\widetilde{\Delta\alpha}
 =
\bigg(
\frac{g}{2} - 1
\bigg)\,I_\gamma
 -
\frac{g}{2}I_{\gamma i},
\label{eq:SpinPrecess.4p}
\end{equation}
where 
\begin{equation}
I_\gamma = \int_0^{\theta}\,\gamma(\theta') d\theta',
\quad\hbox{and}\quad
I_{\gamma i} = \int_0^{\theta}\,\frac{d\theta'}{\gamma(\theta')}.
\label{eq:SpinPrecess.4q}
\end{equation}
To account for fringe fields two more terms, 
$\widetilde{\Delta\alpha}^{\rm FF,in}$ and 
$\widetilde{\Delta\alpha}^{\rm FF,out}$, will later be added directly 
to the right hand side of Eq.~(\ref{eq:SpinPrecess.4p}). This treatment
makes the less-than-perfect assumption that the precession axes in the
fringe fields are normal to the bend plane, which makes it valid to
simply sum the bend field and fringe field precessions.

\paragraph{\bf Spin tracking through fringe fields.}
So far, as a particle enters or exits a bend element, 
its potential
energy has been treated as changing discontinuously with 
its kinetic energy changing correspondingly. For spin tracking,
because there is significant excess spin precession in fringe 
fields we have to treat this region more carefully. Instead of 
treating the potential as discontinuous, we now assume the change 
occurs over a longitudinal distance $\Delta z^{\rm FF}$ which, for 
estimation purposes, we take equal to the separation 
distance (symbol $gap$) between the electrodes; $\Delta z^{\rm FF}=gap$. 
(For the ``protonium'' model, introduced later for comparison with
with analytic SCT calculations, the drift lengths
are taken to be negligably small, making the fringe field spin 
precession negligible.) For orbit tracking the fringe field region 
is assumed to be short enough to be treated as ``thin''. That is, 
any change in the particle's radial offset
occuring in range $\Delta z^{\rm FF}$ is to be neglected and the
integrated deflection applied at the center (i.e. the
edge of the bend). As a result 
the curve $x(z)$ is continuous, but its slope $dx/dz$ is 
discontinuous. Entrance transitions 
from outside a bend to inside are described first.

Inside the bending element
the increase in potential energy from orbit
centerline to radial position $x$ is $e\Delta V(x)$. 
As synchroton oscillations move the particle radially in
and out, the sign of $\Delta V(x)$, just inside the
bend edge oscillates between negative and positive values, 
and the sign of the deviation from the magic velocity 
oscillates correspondingly. This will tend to
average away the spin run-out occurring in the fringe field
region over times long compared to
the synchrotron period. In the long run it is the deviation 
from zero of this average that has to be determined. This can be
done by pure numerical tracking or theoretically or, 
possibly, by a theoretically-weighted averaging of the 
numerical tracking data.

Once one is able to determine the spin decoherence the task will
shift to designing sextupole distributions capable of increasing
the spin coherence time SCT. Our approach will be to study
the effectiveness of such schemes before attempting to improve
the precision of our fringe field treatment.

The deflection angle $\theta^{(FF)}$ of the design orbit in
the fringe field at one such edge is approximately
\begin{equation}
\theta^{(FF)}
 \approx
\frac{1}{2}\,
\frac{\Delta z^{\rm FF}}{r_0}
\quad
\Big(
\overset{\rm e.g.}{\ =\ }
0.5\times0.03/40
 =
0.375\times10^{-3}
\Big);
\label{eq:FF.1}
\end{equation}
this is half of the deflection occurring in advancing a
distance $gap$ in the interior of the bend. (The angle $\theta^{(FF)}$ 
is implicitly assumed to be positive, irrespective of whether the
orbit is clockwise or counter-clockwise.)
Consider a particle approaching the fringe field region
at radial displacement $x$. At the longitudinal center of the fringe field
region the kinetic energy of this particle deviates
from its ``proper'' (i.e. fully-inside value at radial displacement $x$)
by the amount
\begin{equation}
\Delta\gamma^{(FF)}(x)
 \approx
\frac{1}{2}\,
\frac{\Delta V(x)}{m_pc^2/e}
 \approx
\frac{1}{2}\,
\frac{E\,x}{m_pc^2/e},
\label{eq:FF.2}
\end{equation}
where $\Delta V_{\rm tot}$ is the total voltage 
increase from inner electrode to outer electrode.
(The electric 
field points radially inward in order for
positive particles to bend toward negative $x$
but, by convention, $E$ is positive.)
Here, for simplicity, we are neglecting the
fact that the actual electric field will have
more complicated $x$-dependence depending,
for example, on the value of the field index $m$. 
Our assumed fringe field spatial dependence is also 
simplistic.

The leading effect of passage through a bend region
with $\gamma$ deviation from magic $\Delta\gamma$,
is a rate of change of spin angle $\alpha$ per
unit deflection angle $\theta$ given by
\begin{equation}
\frac{d\alpha}{d\theta}
 \approx
\bigg(\frac{g}{2} - 1 + \frac{g/2}{\gamma_0^2}\bigg)\,\Delta\gamma
\quad
\Big( \overset{\rm for\ proton}{\quad=\quad}
3.586\,\Delta\gamma.
\Big)
\label{eq:FFsinglePass.1}
\end{equation}
Combining equations, the excess angular advance occurring
while entering the bend at displacement $x$ is 
\begin{equation}
\boxed{
\widetilde{\Delta\alpha}^{\rm FF}
 =
+\bigg(\frac{g}{2} - 1 + \frac{g/2}{\gamma_0^2}\bigg)\,
\frac{1}{2}\,
\frac{E\,x}{m_pc^2/e}\,
\theta^{FF}.
\label{eq:FFsinglePass.1p}
}
\end{equation}
({\bf Aside:} it may be appropriate to keep another term in
expansion~(\ref{eq:FFsinglePass.1}) in order to include
the fact that dispersion introduces a correlation between
$\gamma$ and $x$ which, after averaging, leaves a finite
precession, even if $<x>$ vanishes
in Eq.~(\ref{eq:FFsinglePass.1p}) .)

In our initial treatment of this edge effect we are assuming
this precession lies in exactly the same plane as the
orbit plane of the particle in the bend element, justifying
the notation $\widetilde{\Delta\alpha}^{\rm FF}$
Entrance (and, later, exit) values can then simply be added 
to the main precession through the bend element.
Meanwhile, in the fringe field region the advance of the tangent 
to the orbit is $\theta^{(FF)}$ as given by Eq.~(\ref{eq:FF.1}).
The + sign on the rhs of Eq.~(\ref{eq:FFsinglePass.1p}) 
reflects the fact that,
for a particle displaced radially outward, the particle momentum
is completing some of its rotation in the fringe field where 
its magnitude is more positive than in the bend interior.

Though the fringe field precession occurs continuously
over the range $gap$ it is applied discontinuously at
the bend edge. This is consistent with our hard edge treatment
of the particle's momentum evolution.
Because $\widetilde{\alpha}$ is measured relative to the orbit direction,
Eq.~(\ref{eq:FFsinglePass.1p}) gives the spin angle precession
over and above the advance of the tangent to the orbit.

The fact that spin and momentum angular advances do not 
match has come about because the particle has bent appreciably 
while its speed deviates from the magic value. On exiting the 
bend element the particle bends similarly while its 
$\gamma$ deviation is given by the same formula~(\ref{eq:FF.2}).
Eq.~(\ref{eq:FFsinglePass.1p}) therefore applies to both entrance 
and exit. Unfortunately this means that excess input precession and 
excess output precession combine constructively rather than tending 
to cancel (as edge focusing commonly does.)

The largest magnitude $\Delta\gamma^{(FF)}$ can have is
\begin{align}
|\Delta\gamma^{(FF)}_{\rm max}|
 &=
\frac{1}{4}\,
\frac{E\,gap}{m_pc^2/e}
\quad
\Big(
\overset{\rm e.g.}{\ =\ }
\frac{1}{4}\,
\frac{(10.5\times10^6)\times0.03}{0.938\times10^9}\, \notag\\
 &=
0.84\times10^{-4}.
\Big)
\label{eq:FF.3m}
\end{align}
For a particle with magic velocity skimming the outer electrode,
where the effect is maximum,
the angular runout is given by 
\begin{align}
|\Delta\alpha^{(FF)}_{\rm max}|
 &\approx
\bigg(\frac{g}{2} - 1 + \frac{g/2}{\gamma_0^2}\bigg)\,
\Delta\gamma_{\rm max}\,
\Delta\theta^{(FF)}         \notag\\
 &=
3.586
\times 
(0.84\times10^{-4})
\times
(0.375\times10^{-3})      \notag\\
 &=                      
1.13\times10^{-7}\,{\rm radians/edge}.
\label{eq:FFsinglePass.2}
\end{align}
With perhaps 50 edges in the lattice, and revolution frequency of 
about 1\,MHz, the maximum spin runout will be about one
revolution per second. This vastly exaggerates the spin
decoherence, of course, because it does not account for
the averaging effect of synchrotron oscillations.
A challenge for lattice design is
to perfect the synchrotron oscillation averaging to zero. 

\paragraph{\bf Spin tracking through thin elements.}
Most of the elements in a storage ring cause
spin precession which approximately conserves the
vertical component of spin $s_y{\bf\hat y}$. The leading
exceptions to this in a proton EDM storage rings are the
quadrupoles (either focusing or defocusing)  
present in the lattice to keep $\beta_y$ much smaller
than $\beta_x$. 
Particles having non-vanishing vertical betatron amplitude are
deflected vertically which causes $s_y{\bf\hat y}$ to
precess. There is a very strong tendency for this precession 
to cancel in subsequent quadrupoles and, therefore, probably 
not contribute significantly to spin decoherence. 
Nevertheless it is essential for this precession to be modeled
correctly to avoid spurious EDM-mimicking precession.
All quadrupoles and sextupoles in the lattice
cause similar precession to at least some degree. 

In ETEAPOT the only thick elements are bends and drifts. 
Spin evolution through them has already been discussed.
All other elements are treated as thin element 
(position dependent) kicks. Copying from the treatment 
in bends, spin evolution through a thin element is 
described by
\begin{equation}
\boxed{
|\widetilde{\Delta\alpha}|
 \approx
\bigg(\frac{g}{2} - 1 + \frac{g/2}{\gamma_0^2}\bigg)\,
\Delta\gamma\,
\widetilde{\Delta\theta}
\label{eq:ThinEl.1m}
}
\end{equation}
where the deflection$\widetilde{\Delta\theta}$ 
is (for now) defined to be positive 
and $\widetilde{\Delta\alpha}$ is the resulting angular 
deviation of the ``bend'' plane spin coordinate relative to 
the orbit; here ``bend'' is in quotes to emphasize the fact
that the bend plane is not, in general, horizontal. By far
the most important instance of this is the deflection of
vertically offset orbits in lattice quadrupoles.
For a particle with transverse position $(x,y)$, the 
magnitude of the angular deviation 
$\widetilde{\Delta\theta}$ in a quadrupole of strength $q$
is given by
\begin{equation}
\widetilde{\Delta\theta} = |q|\sqrt{x^2+y^2},
\label{eq:ThinEl.1p}
\end{equation}
where $q$ is the inverse focal length of the quadrupole.
The absolute value sign in this equation eventually has
to be removed; it is included here so that the discussion
of signs can be deferred. 

The magnitude of the angular deflections in sextupoles and 
higher order multipoles are also functions only of the
combination $r=\sqrt{x^2+y^2}$. Eq.~(\ref{eq:ThinEl.1p}) 
generalizes to 
\begin{equation}
\widetilde{\Delta\theta}_{\rm quad} = |q|r,\quad
\widetilde{\Delta\theta}_{\rm sext} = \frac{|S|}{2}\,r,\quad
\widetilde{\Delta\theta}_{\rm oct} = \frac{|O|}{6}\,r,
\label{eq:QuadSextOct.1}
\end{equation}
where $S$ is the conventionally defined sextupole strength and 
$O$ is the conventionally defined octupole 
strength. 
\footnote{The element strengths appearing in an SXF 
lattice description file include the integer factors.
That is, the quad parameter is $b_1=q$, the sextupole parameter 
is $b_2=S/2$, the octupole parameter is $b_3=S/6$, and so on.}

A significant complication concerns the sign of
$\widetilde{\Delta\alpha}$. For a particle with no vertical
displacement there is no ambiguity, since the bend plane
in the quadrupole is the same as the overall (horizontal)
lattice design plane. In this case, with $y=0$,
Eq.~(\ref{eq:ThinEl.1m}) can be made more explicit;
\begin{equation}
\widetilde{\Delta\alpha}_h
 \approx
\bigg(\frac{g}{2} - 1 + \frac{g/2}{\gamma_0^2}\bigg)\,
\Delta\gamma\,
q\,x.
\label{eq:ThinEl.1q}
\end{equation}
where, by convention, a horizontal \emph{focusing} quad has $q>0$. 
The sign in Eq.~(\ref{eq:ThinEl.1q}) reflects the fact that, for 
$x>0$ and $q>0$, the quadrupole ``helps'' by 
bending the momentum in the same sense as the bending elements.
This formula makes it clear that reversing the sign of $q$
reverses the sign of $\widetilde{\Delta\alpha}$.

For obtaining the proper sign for general multipoles it is 
necessary to handle consistently
the transformation from laboratory to bend frame spin
coordinates, which is why the sign issue has been deferred.

For a particle incident at 
$(x_0,y_0)$ the equation of the line of intersection
of the deflection plane with the transverse plane is
\begin{equation}
y = y_0 - \frac{y_0}{x_0}\,(y-y_0),
\label{eq:ThinEl.2}
\end{equation}
In a quadrupole the roll-angle of the deflection plane
(with counter-clockwise roll taken as positive)
is $\phi_0=\tan^{-1}(y_0/x_0)$,
irrespective of quadrant and whether the quadrupole is
focusing or defocusing. However the inverse tangent function
is, itself, multiple valued. To make it single valued one can
determine $\phi_0$ using $\phi_0={\rm arctan2}(qy_0,qx_0)$.

Along with Eq.~(\ref{eq:ThinEl.1m}), this establishes
both the sign and magnitude of $\widetilde{\Delta\alpha}$,
while preserving the sign reversal when the sign of $q$ reverses.
Including also sextupoles, octupoles and other multipoles,
one obtains
\begin{align}
\phi_{\rm 0,quad} &= 1\,{\rm arctan2}(qy_0,qx_0), \notag\\
\phi_{\rm 0,sext} &= 2\,{\rm arctan2}(Sy_0,Sx_0), \notag\\
\phi_{\rm 0,oct} &= 3\,{\rm arctan2}(Oy_0,Ox_0).
\label{eq:QuadSextOct.2}
\end{align}
With roll angles of the bend plane determined this way, the
following formulas apply to all multipoles.

The required spin rotation matrix is
\begin{align}
S_m 
&=
{\bf I} +
2\sin(\widetilde{\Delta\alpha}/2)
\begin{pmatrix} c_{11} & c_{12} & c_{13} \\
                c_{21} & c_{22} & c_{23} \\
                c_{31} & c_{32} & c_{33}
\end{pmatrix}       
\notag\\
&=
{\bf R_m}(-\phi_0)
\begin{pmatrix} 
 \cos\widetilde{\Delta\alpha}  &  0  & -\sin\widetilde{\Delta\alpha}  \\
           0                   &  1  &              0                 \\
 \sin\widetilde{\Delta\alpha}  &  0  &  \cos\widetilde{\Delta\alpha}  
\end{pmatrix}
{\bf R_m}(\phi_0)\notag\\
\hbox{where}\quad &
{\bf R_m}(\phi)
  =
\begin{pmatrix} 
  \cos\phi & -\sin\phi  &  0 \\
  \sin\phi &  \cos\phi  &  0 \\
    0      &     0      &  1
\end{pmatrix}.
\label{eq:ThinEl.4}
\end{align}
To avoid serious loss of numerical precision, it is essential to 
use trigonometric identities to express the matrix product exactly 
and explicitly in this way. This utilizes the fact that the 
dominant, identity matrix part of the central matrix commutes with 
the outer matrices, whose product is ${\bf I}$.
The remaining, exact matrix elements are are explicitly small. 
\begin{align}
c_{11} &= -\cos(\phi)^2\sin(\widetilde{\Delta\alpha}/2),        \notag\\
c_{12} &=  \cos(\phi)\sin(\widetilde{\Delta\alpha}/2)\sin(\phi),\notag\\
c_{13} &= -\cos(\phi)\cos(\widetilde{\Delta\alpha}/2),          \notag\\
c_{21} &=  \cos(\phi)\sin(\widetilde{\Delta\alpha}/2)\sin(\phi),\notag\\
c_{22} &= -\sin(\phi)^2\sin(\widetilde{\Delta\alpha}/2),        \notag\\
c_{23} &=  \sin(\phi)\cos(\widetilde{\Delta\alpha}/2),        \notag\\
c_{31} &=  \cos(\phi)\cos(\widetilde{\Delta\alpha}/2),        \notag\\
c_{32} &= -\sin(\phi)\cos(\widetilde{\Delta\alpha}/2),        \notag\\
c_{33} &= -\sin(\widetilde{\Delta\alpha}/2)
\label{eq:ThinEl.4pp}
\end{align}
Since the common factor $2\sin(\widetilde{\Delta\alpha}/2)$ is tiny, 
all matrix elements are small even for angles $\phi$ of order $\pi$. 
Multiplying this matrix on the right by $(s_x,s_y,s_z)^T$
produces deviations $(\Delta s_x,\Delta s_y,\Delta s_z)$
which, added to $(s_x,s_y,s_z)$, give the output spin coordinates.

\end{document}